\documentclass[10pt, aps, prd, superscriptaddress, nofootinbib, showpacs, twocolumn]{revtex4-2}
\pdfoutput=1
\usepackage{amsmath,amssymb}
\usepackage{epsfig,graphicx}
\usepackage{color}
\usepackage{hyperref}

\numberwithin{equation}{section}

\newcommand{\calE}{\mathcal E}

\begin{document}

\title{Notes on conformal anomaly, nonlocal effective action and the metamorphosis of the running scale}

\author{A. O. Barvinsky}
\email{barvin@td.lpi.ru}
\affiliation{Theory Department, Lebedev Physics Institute, Leninsky Prospect 53, Moscow 119991, Russia}
\affiliation{Institute for Theoretical and Mathematical Physics, Moscow State University, Leninskie Gory, GSP-1, Moscow, 119991, Russia}

\author{W. Wachowski}
\email{vladvakh@gmail.com}
\affiliation{Theory Department, Lebedev Physics Institute, Leninsky Prospect 53, Moscow 119991, Russia}

\begin{abstract}
We discuss the structure of nonlocal effective action generating the conformal anomaly in classically Weyl invariant theories in curved spacetime. By the procedure of conformal gauge fixing, selecting the metric representative on a conformal group orbit, we split the renormalized effective action into anomalous and Weyl invariant parts. A wide family of thus obtained anomalous actions is shown to include two special cases of Riegert--Fradkin--Tseytlin and Fradkin--Vilkovisky actions. Both actions are shown to be contained in the first three orders of the curvature expansion for a generic one-loop effective action obtained by covariant perturbation theory. The complementary Weyl invariant part of the action is given by the ``conformization'' of the full effective action---restricting its argument to the conformally invariant representative of the orbit of the conformal group. This is likely to resolve a long-standing debate between the proponents of the Riegert action and adherents of the perturbation expansion for the effective action with typical nonlocal logarithmic form factors. We derive the relation between quantum stress tensors on conformally related metric backgrounds, which generalizes the known Brown-Cassidy equation to the case of nonzero Weyl tensor, and discuss applications of this relation in the cosmological model driven by conformal field theory. We also discuss the issue of renormalization group running for the cosmological and gravitational coupling constants and show that it exhibits a kind of a metamorphosis to the nonlocal form factors of the so-called partners of the cosmological and Einstein terms---nonlocal curvature squared terms of the effective action.
\end{abstract}

\maketitle
\tableofcontents

\vspace{1cm}
\begin{flushright} \it To the memory of Stanley Deser \end{flushright}

\section{Introduction}

The status of local Weyl anomalies is widely considered to be fully settled in current literature. However, the issue of their relevance to concrete physical effects, as opposed to a mere criterion of consistency at the quantum level of the classically Weyl invariant theories, often remains a subject of the debate. The manifestation of the conformal anomaly in physical applications usually occurs within the effective action formalism, and there is extending over years debate on the structure of this action, taking place between the pioneers of the conformal anomaly and adherents of perturbation theory. The nature of this debate consists in a seemingly contradictory difference between the known expression for the anomaly action and the form of the nonlocal effective action obtained by Feynman diagrammatic technique.

As is well known, the one-loop conformal anomaly for classically Weyl invariant 4-dimensional theory having in Euclidean curved spacetime the covariantly renormalized effective action $\varGamma[\,g_{\mu\nu}]$ reads as \cite{Capper-Duff0,Capper-Duff1,DDI,Duff:1977ay,Birrell-Davies,Duffanomaly,Deser96}
\begin{align}
    \hspace{-0.2cm}&\langle\, T^\mu_\mu\,\rangle \equiv \frac{2\,g_{\mu\nu}}{\sqrt{g}} \frac{\delta
    \varGamma}{\delta g_{\mu\nu}} = \frac1{16\pi^2} \big(\alpha C^2 + \beta E +\gamma\Box R\big),         \label{anomaly}\\
    &\;E= R_{\mu\nu\alpha\gamma}R^{\mu\nu\alpha\gamma} - 4R_{\mu\nu}R^{\mu\nu} + R^2,
\end{align}
where $\sqrt{g}E$ denotes the Gauss--Bonnet density, $C_{\mu\nu\alpha\beta}$ is the Weyl tensor, $C^2 = C_{\mu\nu\alpha\beta}C^{\mu\nu\alpha\beta}$, and $\alpha$, $\beta$ and $\gamma$ are the numerical coefficients depending on the spin of the quantum field.\footnote{We work in Euclidean signature spacetime, and our notations are $R^\alpha{}_{\beta\mu\nu}=\partial_\mu\varGamma^\alpha_{\nu\beta} - \cdots$, $R_{\mu\nu}=R^\alpha{}_{\mu\alpha\nu}$, $\Box=g^{\mu\nu}\nabla_\mu\nabla_\nu$. For simplicity we do not include in the anomaly the contribution $F_{\mu\nu}^2$ of the vector gauge field and $\varphi^4$-contribution of the self-interacting conformal scalar field.} The anomalous action $\varGamma_A[\,g_{\mu\nu}]$ generating this anomaly was first derived in the nonlocal form by Riegert \cite{Riegert} and by Fradkin and Tseytlin \cite{FT} in the local form of the conformal Wess-Zumino action involving an auxiliary scalar field---the dilaton responsible for intetwining two conformally related metrics. The nonlocal form of the Riegert--Fradkin--Tseytlin (RFT) action reads as
\begin{align}
    \varGamma_A[\,g\,] &= \frac{1}{64\pi^2}\int d^4x\, \sqrt{g}\,
    \Big(\alpha\, C^2 + \frac\beta2 \calE_4\Big) \frac1\Delta_4\calE_4 \nonumber \\
    &-\frac{1}{32\pi^2} \Big(\frac{\gamma}{6}+\frac\beta{9}\Big)\,\int d^4x\,\sqrt{g}R^2,   \label{RFT}
\end{align}
where
    \begin{equation}
    {\cal E}_4 \equiv E - \tfrac{2}{3}\Box R, \label{calE}
    \end{equation}
$\Delta_4$ denotes the so-called Paneitz operator \cite{Paneitz}
    \begin{equation} \label{Paneitz}
    \Delta_4 = \Box^2 + 2R^{\mu\nu}\nabla_{\mu}\nabla_{\nu} - \frac{2}{3} R\,\Box + \frac{1}{3}(\nabla^{\mu}R)\,\nabla_{\mu}
    \end{equation}
and $1/\Delta_4$ implies its inverse---the notation for the operation of acting by its Green's function ${\cal G}(x,y)$ on a generic test function $\psi(y)$, $\Delta_4{\cal G}(x,y)= \delta(x,y)$, $\tfrac1\Delta_4 \psi(x)= \int d^4y\,{\cal G}(x,y)\,\psi(y)$.

Some time after the invention of the RFT action the attention to it was drawn by Antoniadis, Mazur and Mottola due to several applications in gravity theory \cite{AM,AMM}, but this caused a serious criticism \cite{Deser-Schwimmer} of the expression \eqref{RFT} in view of its drastic structural difference from the renormalized effective action built within perturbation theory in powers of spacetime curvature. This expansion begins with \cite{DDI}
    \begin{align}
    \varGamma_{\rm ren} &= \frac1{32\pi^2}
    \int dx\,\sqrt{g} \Big[-\alpha\,
    C_{\mu\nu\alpha\beta}
    \ln\Big(\!-\frac{\Box}{\mu^2}\Big)C^{\mu\nu\alpha\beta} \nonumber \\
    &-\frac\gamma6 R^2\,\Big] + O(\Re^3),            \label{Gamma_quad}
    \end{align}
$\Re$ collectively denoting here the Riemann, Ricci and scalar curvature, and does not at all resemble the form of \eqref{RFT}. This criticism was maintained by objections against short distance behavior of stress tensor correlation functions generated by the RFT action, which were shown to contradict the conformal Ward identities for these correlator \cite{Erdmenger-Osborn}. Another criticism was associated with the objections against the double pole structure of the Green's function of the operator \eqref{Paneitz}, $\sim 1/\Box^2$ \cite{Deser}. Although these objections were disclaimed in \cite{Coriano-Maglio-Mottola} by explicit calculations of $\langle TTT\rangle$-correlators, the question might still be hovering unsettled in the literature \cite{Duff_on_Freund}.

The goal of this paper will be to discuss the status of the effective action responsible for the generation of the Weyl anomaly. To begin with we will focus on a wide variety of nonlocal anomalous actions by including the RFT action in their functional family. The idea of this construction is similar to gauge fixing applied to the ambiguity of the conformal split of the metric argument of the action functional, which was suggested rather long ago in \cite{BMZ}. The resulting class of anomaly actions will be parameterized by the conformal gauge selecting the representative on the orbit of the local conformal group. We will explicitly demonstrate that the difference between the members of this class is a Weyl invariant functional---a point of departure between various suggestions for the anomalous action. Two particular gauges will be considered, one of them exactly corresponding to the RFT action \eqref{RFT} and another associated with the Weyl invariant nonlocal rescaling of the metric field suggested by Fradkin and Vilkovisky.  This rescaling, which is directly applicable in asymptotically flat spacetimes, was designed as a remedy against the trace anomaly \cite{FV}---the analogue of the Yamabe problem of a local Weyl transformation to the metric with a vanishing scalar curvature.

Then we show how the Fradkin--Vilkovisky version of the anomaly action arises in the first three orders of the covariant curvature expansion for a generic one-loop effective action. We discuss the associated mechanism of partial summation of scalar curvature terms of this expansion \cite{MVZ} along with the double pole problem for the Green function of the Paneitz operator \eqref{Paneitz}.

Lack of uniqueness of the anomaly action defined only up to a Weyl invariant functional raises, of course, the question of its incompleteness in concrete applications. This also poses the question of whether the RFT action or its modifications within the above class provides an optimal description of the physical problem in question. For example, it is well known that in two dimensions the stress tensor trace anomaly and the associated nonlocal Polyakov action are fully responsible for the Hawking radiation of the two dimensional black holes \cite{Christensen-Fulling}. On the contrary, in higher dimensions the anomaly action is insufficient to describe this phenomenon. Still there is a strong belief \cite{AM,AMM,Coriano-Maglio-Mottola} that at distances of the horizon scale gravity theory is essentially modified due to large infrared effects of the conformal mode described by the action \eqref{RFT}. These effects might dominate macroscopic physics at such scales, like for instance the near black hole horizon behavior of quantum stress tensor \cite{PhysRevD.74.064004}, the contribution to the scalar sector of gravitational waves \cite{Mottola:2016mpl} or dynamical vacuum energy in effective theory of gravity \cite{Mottola:2022tcn}. Though it is not entirely clear how complete is the setup in these problems, there are physical situations when the conformal mode really runs the whole show, and we consider as a direct application of \eqref{RFT} two examples of such a situation. These are the calculation of the metric stress tensor in a generic conformally flat spacetime \cite{Brown-Cassidy} and the Friedmann metric cosmology driven by the trace anomaly of conformal invariant fields \cite{FHH}, the latter playing important role in the model of initial conditions for inflationary cosmology \cite{slih,why}.

A related issue in the problem of nonlocal effective action is the question of renormalization group (RG) running of the cosmological and gravitational constants. Though the issue of running scale and its relation to the cosmological constant problem has already become a byword in current literature, it becomes increasingly clearer that this running should not be interpreted in the usual sense of RG theory \cite{Donoghue_grav_const,Donoghue}. The notion of ``scale'' is so ambiguous in physics that its running nature actually looses universality when addressing various physical setups, like for example associating cosmological inflation with RG running \cite{Woodard}. Serious arguments against running nature of the cosmological and gravitational couplings in \cite{Donoghue_grav_const,Donoghue} have led to the notion of cosmological constant partners \cite{Donoghue_partners} interpreted in \cite{Gorbar-Shapiro_running} in terms of separation of scales or decoupling of heavy modes \cite{Appelquist:1974tg,Gorbar-Shapiro_decoupling}. Still, it is customary to have nontrivial solutions of RG equations in renormalizable gravity models \cite{Tseytlin_quadratic,Avramidy-B} with running scale dependent $\varLambda$ and $G$. Therefore a natural question arises how these solutions have to be interpreted when the tadpole structure of the covariant cosmological and Einstein terms preclude them from their actual dependence on the momentum \cite{Donoghue}.

So one of the goals of this paper is an attempt to clarify this issue within a special version of the notion of the ``scale''. Looking forward to the final conclusion, we might formulate the suggestion for the notions of running $\varLambda$ and $G$ couplings as their conversion or metamorphosis into their nonlocal partners similar to those introduced by J. Donoghue in \cite{Donoghue_partners}. Within perturbation scheme the cosmological and Einstein terms start manifesting themselves as nonlocal curvature squared terms very different from their original form.

The paper is organized as follows. In Sect.~\ref{ConfGaugeSect} we decompose the quantum effective action into anomalous and Weyl invariant parts by imposing the conformal gauge for the choice of the representative on the orbit of the conformal group. This allows one to build the whole class of nonlocal anomalous actions, functionally parameterized by the choice of this gauge and including the RFT action \eqref{RFT} and the Fradkin--Vilkovisky action suggested in \cite{MVZ}. Sect.~\ref{CurvExpansSect} contains the discussion of the covariant curvature expansion of \cite{CPTI,CPTII,CPTIII} and the way how it contains the anomalous action in the lowest orders of this expansion. In particular, it is shown that the Fradkin--Vilkovisky version of this action performs a resummation of the covariant curvature series in powers of the Ricci scalar \cite{MVZ}. In Sect.~\ref{StressTensorSect} we give a direct and, apparently, not very well known derivation from the RFT action of the vacuum stress-tensor behavior at the orbit of the conformal group---a good example of direct applicability of \eqref{RFT}. Here we also comment on the application of the anomalous conformal Wess-Zumino action to the {\em a}-theorem \cite{Jack-Osborn,KomargodskiSchwimmer} and present the generalization of the Brown--Cassidy formula \cite{Brown-Cassidy} for the stress tensor to the case of a nonzero Weyl tensor, see Eq.(\ref{BC_gen}). Applications of the anomaly action in conformally flat spacetime are presented in Sect.~\ref{ConfFlatSect}. It is shown how this action underlies the construction of the inflation scenario starting from the cosmological initial state in the form of the mircocanonical density matrix \cite{slih, why, slih_R^2, hill-top, CHS}, recently reviewed in \cite{SLIH_review}. Important feature of this application is the value of the Casimir vacuum energy which is also determined by the coefficients of the anomalous trace \eqref{anomaly} \cite{Dowker-Critchley, Mamaev-Mostepanenko-Starobinsky, Ford, Candelas-Dowker}.

In Sect.6 we discuss the problem of scale dependence of the gravitational and cosmological constants related to the ideas of \cite{Donoghue_grav_const, Donoghue,Donoghue_partners} and \cite{covnonloc, covlongdist}. Here we show that in the UV regime the RG analysis of the cosmological and Einstein terms strongly points out to the conversion of their scale dependence into the nonlocal form factors of their UV partners represented by curvature squared terms with dimensionless nonlocal coefficients. We call this phenomenon a metamorphosis of the running scale, which we derive by using a special scaling operator. In IR domain the same analysis leads to the low energy partners depending on mass scale of the theory. These nonlocal partners were suggested in \cite{Donoghue_partners} by J. Donoghue for the cosmological constant term and blueprinted for the Einstein term in \cite{covnonloc, covlongdist} in the form of the long distance modification of Einstein gravity.

In the concluding section we briefly recapitulate the above observations and dwell on related potential problems and applications. We start by discussing the role of Weyl anomaly in the problem of cosmological initial conditions for the inflation scenario driven by a conformal field theory \cite{SLIH_review}. This scenario motivates introduction of numerous conformal higher spin (CHS) fields whose Weyl anomaly is generated only in the one-loop approximation and, thus, acquires a kind of nonperturbative status. Then we discuss the uniqueness for the nonlocal scaling operator used for the derivation of the above metamorphosis phenomenon. In particular, we show that in the curvature squared terms of the action it is nearly uniquely determined due to general covariance of the theory, though in Lorentz symmetry violating models like Ho\v{r}ava gravity \cite{Horava} it may be rather ambiguous.

\section{Conformal gauge fixing} \label{ConfGaugeSect}

The splitting of the renormalized effective action of a classically conformally invariant theory into the anomaly part $\varGamma_A$ generating the trace anomaly \eqref{anomaly} and the Weyl invariant part $\varGamma^{\rm conf}$, $g_{\mu\nu}\delta \varGamma^{\rm conf}/\delta g_{\mu\nu}=0$,
    \begin{equation}
    \varGamma_{\rm ren}=\varGamma_A+\varGamma^{\rm conf},
    \end{equation}
is obviously not unique and admits the freedom
    \begin{equation}
    \varGamma_A\to \varGamma_A+W^{\rm conf},\qquad
    \varGamma^{\rm conf}\to\varGamma^{\rm conf}-W^{\rm conf}, \label{shifting}
    \end{equation}
with an arbitrary conformally invariant functional $W^{\rm conf}$,
\begin{equation}
g_{\mu\nu}\frac{\delta W^{\rm conf}}{\delta g_{\mu\nu}}=0.
\end{equation}
The freedom in the choice of $W^{\rm conf}[\,g_{\mu\nu}]$ arises as a functional integration constant for the first order variational equation that can be written down for $\varGamma_A[\,g_{\mu\nu}]$ or for the renormalized effective action $\varGamma[\,g_{\mu\nu}]\equiv\varGamma_{\rm ren}[\,g_{\mu\nu}]$. At the orbit of the conformal group passing through the metric $g_{\mu\nu}$---the argument of the effective action---and parameterized by the local conformal parameter $\sigma=\sigma(x)$,
    \begin{equation} \label{orbit}
    g_{\mu\nu} = e^{2\sigma} \bar g_{\mu\nu},
    \end{equation}
the renormalized action $\varGamma_{\rm ren}[\,e^\sigma\bar g\,]$ satisfies the equation
    \begin{equation} \label{orbiteq}
    \frac{\delta \varGamma_{\rm ren}[e^{2\sigma}\bar g]}{\delta\sigma} = \frac{\sqrt{g}}{16\pi^2} \big(\alpha C^2 + \beta E +\gamma \Box R\big) \Big|_{g_{\mu\nu} = e^\sigma\bar g_{\mu\nu}},
    \end{equation}
which can be integrated to give conformal Wess-Zumino action \cite{FT}
    \begin{align}
    &\Delta\varGamma[\,\bar g, \sigma\,] \equiv \varGamma_{\rm ren}[\,g\,] - \varGamma_{\rm ren}[\,\bar g\,] \nonumber \\
&\qquad= \frac{1}{16\pi^2}\int d^4x\, \sqrt{\bar g} \left\{ \Big[\alpha\bar C^2 + \beta\bar\calE_4\Big] \sigma + 2\beta\sigma\bar\Delta_4\sigma\right\} \nonumber \\
  &\qquad- \frac{1}{32\pi^2} \Big(\frac\gamma6 + \frac{\beta}9\Big)\,\int d^4x\,\Big(\sqrt{g}R^2 - \sqrt{\bar g}\bar R^2\Big),          \label{RTF}
    \end{align}
where the two metrics $g_{\mu\nu}$ and $\bar g_{\mu\nu}$ are related by the equation \eqref{orbit}, all barred quantities are built in terms of $\bar g_{\mu\nu}$ and $\bar\Delta_4$ is the barred version of the fourth-order Paneitz operator \eqref{Paneitz}. This expression $\varGamma_{\rm ren}-\bar\varGamma_{\rm ren}=\varGamma_A -\bar\varGamma_A$ can also be rewritten in the other form
    \begin{align}
    &\varGamma_A[\,g\,]-\varGamma_A[\,\bar g\,] \nonumber\\
&\qquad= \frac{1}{16\pi^2}\int d^4x \sqrt{g}\,\left\{\,
    \Big[\,\alpha\, C^2 + \beta\calE_4\Big]\, \sigma
    -2\beta\,\sigma\Delta_4\sigma\,\right\}\nonumber\\
    &\qquad- \frac{1}{32\pi^2} \Big(\frac\gamma6 + \frac{\beta}{9}\Big)\,\int d^4x\,\Big(\sqrt{g}R^2 - \sqrt{\bar g}\bar R^2\Big),          \label{RTF1}
    \end{align}
if one takes into account two important properties of the Paneitz operator---Weyl invariance of its densitized form,
    \begin{equation}
    \sqrt{\bar g}\,\bar\Delta_4 = \sqrt{g}\,\Delta_4, \label{conf_Paneitz}
    \end{equation}
and the finite conformal transformation of ${\cal E}_4$---the Gauss-Bonnet density modified by $\sqrt{g}\,\Box R$ term (\ref{calE}),
    \begin{equation} \label{basic}
    \sqrt{g}\,\calE_4 = \sqrt{\bar g}\,\bar\calE_4 + 4\sqrt{\bar g}\,\bar\Delta_4\sigma. \end{equation}
These two properties are consistent with each other because the last equation should obviously remain valid under the interchange of $g_{\mu\nu}$ and $\bar g_{\mu\nu}$ accompanied by flipping the sign of $\sigma$.

There is also the third form of the Wess-Zumino action, which will be given below in Eq.(\ref{minimal}). It exists for a special renormalization converting to zero the coefficient $\gamma$ of the $\Box R$ term in (\ref{anomaly}), and underlies the proof of the so-called $a$-theorem for the monotonic RG flow of the coefficient $a=\beta/16\pi^2$ of the topological term in the trace anomaly \cite{Jack-Osborn,KomargodskiSchwimmer}.

Modulo a nonvanishing conformal anomaly all points on the orbit of the conformal group \eqref{orbit} are physically equivalent, and this typical situation of a broken local gauge invariance can be managed by introducing the gauge condition which uniquely selects $\bar g_{\mu\nu}$ as the representative of the equivalence class of metrics \eqref{orbit}. If we denote this gauge condition as $\chi[\bar g]=0$ then this representative should be uniquely selected by the solution of the equation for the conformal parameter $\sigma$,
    \begin{equation}
    \chi[\,\bar g\,] = \chi[\,g e^{-2\sigma}] = 0,
    \end{equation}
this solution being a functional of the metric $\varSigma_\chi[\,\bar g\,]$, labelled by the gauge symbol $\chi$,
    \begin{equation}
    \sigma = \varSigma_\chi[\,\bar g\,].
    \end{equation}
The representative of the conformal orbit $\bar g_{\mu\nu}[g]$ as a functional of a given metric $g_{\mu\nu}$ (through which the orbit is passing) becomes Weyl invariant,
    \begin{equation}
    \bar g_{\mu\nu}[\,g\,]\equiv g_{\mu\nu}\,e^{-2\varSigma_\chi[\,\bar g\,]},
    \quad
    g_{\alpha\beta}\frac{\delta
    \bar g_{\mu\nu}[\,\bar g\,]}{\delta g_{\alpha\beta}} = 0,  \label{conf_inv}
    \end{equation}
because under any local Weyl rescaling $g_{\mu\nu}\to e^{2\sigma} g_{\mu\nu}$ the conformal parameter transforms as $\varSigma_\chi[g]\to
\varSigma_\chi[g] + \sigma$ in view of the identity $\chi\big[g e^{-\varSigma_\chi[g]}\big] \equiv 0$, so that
    \begin{equation}
    \delta_\sigma\varSigma_\chi[\,\bar g\,]=\sigma,  \label{deltaSigma}
    \end{equation}
where $\delta_\sigma$ is the operator of the conformal variation
    \begin{equation}
    \delta_\sigma\equiv 2\int d^4x\;\sigma(x)\, g_{\mu\nu}(x)\frac{\delta}{\delta g_{\mu\nu}(x)}.
    \end{equation}
For the uniqueness of such conformal gauge fixing procedure (in spacetime and at least in some finite domain of the space of metrics) the Faddeev--Popov operator $Q_\chi=Q_\chi(x,y)$, corresponding to the gauge $\chi[g]$, $\delta_\omega\chi(x)=\int d^4y\,Q_\chi(x,y)\,\omega(y)$,  should be nondegenerate.

Thus, the terms of \eqref{RTF1}
    \begin{align}
    \hspace{-0.2cm}W^{\rm conf}[\,g\,] = \varGamma_A[\,\bar g\,] + \frac1{32\pi^2}\Big(\frac\gamma6 + \frac{\beta}{9}\Big)\!
    \int d^4x
    \sqrt{\bar g}\,\bar R^2
    \end{align}
taken at $\bar g_{\mu\nu}[\,g\,]$ can be considered as an irrelevant Weyl invariant integration ``constant'', while the rest of the terms can be identified with the anomaly action after the substitution of $\sigma=\varSigma_\chi[\,g\,]$. This set of anomaly actions $\varGamma_A[\,g\,]\equiv\varGamma_\chi[\,g\,]$ parameterized and labelled by conformal gauge conditions $\chi$ reads as
    \begin{align}
    &\hspace{-0.2cm}\varGamma_\chi[\,g\,]= \frac{1}{16\pi^2}\int d^4x\, \sqrt{g}\,\Big\{ \big(\alpha C^2 + \beta\calE_4\big) \varSigma_\chi \nonumber \\
    &\hspace{-0.3cm}\;\;\;\;- 2\beta\varSigma_\chi\Delta_4\varSigma_\chi\Big\} - \frac{1}{32\pi^2}
    \Big(\frac\gamma6 + \frac{\beta}{9}\Big)\!\int d^4x\,\sqrt{g}R^2.              \label{chi_action}
    \end{align}

The difference between various members of this set is, of course, a Weyl invariant functional. For two arbitrary conformal gauges one has
    \begin{align}
    \varGamma_{\chi_1} - \varGamma_{\chi_2} &= \frac1{16\pi^2} \int d^4x\, \sqrt{g}\,
    \big(\varSigma_{\chi_1}-\varSigma_{\chi_2}\big) \nonumber\\
&\times\Big[\alpha\, C^2 +\beta\calE_4 -2\beta\Delta_4\big(\varSigma_{\chi_1}
    +\varSigma_{\chi_2}\big)\Big].      \label{Gamma1-Gamma2}
    \end{align}
Conformal variation of this expression is vanishing, because of the transformation law \eqref{deltaSigma} for $\varSigma_{1,2}$, Weyl invariance of the density $\sqrt{g}\,C^2$ and the relation \eqref{basic} which in the infinitesimal form reads as
    \begin{equation}
    \delta_\sigma \Big[\sqrt{g}\,\calE_4\Big] = 4\sqrt{g}\,\Delta_4\sigma,
    \end{equation}
so that using all the above properties  $\delta_\sigma\big(\varGamma_{\chi_1} -\varGamma_{\chi_2}\big)=0$.

Note that with our definition of the anomaly action \eqref{chi_action} the way it enters the full quantum action can be represented as
    \begin{equation} \label{Gamma_vs_Gamma_chi}
    \hspace{-0.1cm}\varGamma[\,g\,] = \varGamma_\chi[\,g\,] +\varGamma[\,\bar g\,] +\frac1{32\pi^2}\Big(\frac\gamma6 + \frac{\beta}{9}\Big) \int d^4x\sqrt{\bar g}\,\bar R^2,
    \end{equation}
where $\bar g_{\mu\nu}[\,g\,] = e^{-2\varSigma_\chi[\,g\,]}g_{\mu\nu}$

\subsection{Riegert--Fradkin--Tseytlin gauge}

An obvious choice of the conformal gauge associated with the Gauss--Bonnet density and the Branson curvature is the Riegert--Fradkin--Tseytlin gauge
    \begin{equation}
    \chi_{{\vphantom L}_{\rm RFT}}[\,\bar g\,] \equiv \bar\calE_4 = 0.          \label{RFT_gauge}
    \end{equation}
It can be imposed for topologically simple spacetime manifolds with a vanishing bulk part of the Euler characteristics (see Eq.(\ref{Euler_surface}) and footnote \ref{Euler} below), to which in particular belongs asymptotically flat spacetime to be mainly considered throughout the paper. The advantage of this gauge is that it is exactly solvable due to the transformation law for the Branson curvature \eqref{basic}. Applying this gauge and using Eq. \eqref{conf_Paneitz} we obtain a linear equation on $\varSigma_{{\vphantom L}_{\rm RFT}}$ which has a solution in terms of the inverse Paneitz operator
    \begin{equation} \label{varSigmaRFT}
    \varSigma_{{\vphantom L}_{\rm RFT}}=\frac14\,\frac1\Delta_4\calE_4.
    \end{equation}
Formally substituting this expression to \eqref{chi_action} we obtain exactly the RFT action \eqref{RFT}.

This RFT action and the inverse Paneitz operator are well defined and exist in asymptotically flat spacetime under Dirichlet boundary conditions at infinity when treated within perturbation theory in powers of the curvatures whose collection is denoted below as $\Re$. Indeed, in this case
    \begin{equation}
   \frac1\Delta_4=\frac1{\Box^2}+O(\Re),
    \end{equation}
and this operator works well when it is applied to the functions of the Branson curvature type $\sim\calE_4$. Because of the double-pole nature of the operator $1/\Box^2$ its action on generic functions may be badly defined due to infrared divergences, but when the function is represented by the total derivative structure it generates, when acted upon by $1/\Box^2$, well defined multipole expansion valid in four dimensions at spacetime infinity \cite{CPTII}.\footnote{\label{footnote}
As discussed in \cite{CPTII}, the operator $1/\Box^n$ in $D$-dimensional space with $D<2n$ is ill defined unless the functions it acts upon are of the form $\partial_{\alpha_1}...\partial_{\alpha_m}j(x)$, $m=2n-D+1$ with the function $j(x)$ having an asymptotic behavior $j(x)=O(1/|x|^D)$, $|x|\to\infty$. This property can be explained by the fact that in the multipole expansion of $\tfrac1\Box\partial_{\alpha_1}... \partial_{\alpha_m}j(x)$ the first few multipoles vanish, which improves the fall-off properties of the result at infinity and makes possible a repeated action by $1/\Box$.}
But the Gauss--Bonnet density and $\sqrt{g}\,\Box R$ are both locally a total derivative which makes $1/\Delta_4$ well defined in the expression \eqref{varSigmaRFT} for $\varSigma_{{\vphantom L}_{\rm RFT}}$. This in fact implies the invertibility of the Faddeev--Popov operator in this gauge, which up to coefficient coincides with the Paneitz operator, $Q_{{\vphantom L}_{\rm RFT}}=4\Delta_4$, and thus guarantees local uniqueness of conformal gauge fixing procedure.

Moreover, the above observation serves as a repudiation of the harmful role of double poles in the RFT action that was claimed in \cite{Deser}. Absence of infrared dangerous double poles is explicit in the lowest order of the curvature expansion for $\varSigma_{{\vphantom L}_{\rm RFT}}$ which reads
\begin{equation}
    \varSigma_{{\vphantom L}_{\rm RFT}}=-\frac1{6\,\Box}R+O(\Re^2), \label{varSigma0}
    \end{equation}
in view of the fact that the Gauss--Bonnet density is quadratic in the curvature $\sqrt{g}E = O(\Re^2)$. Higher orders of this expansion are also safe because of the total derivative nature of $\sqrt{g}\,E$. Regarding the lowest order quadratic in curvature part, with the above approximation for $\varSigma_{{\vphantom L}_{\rm RFT}}$ it equals
    \begin{equation} \label{RFT_quad}
    \varGamma_{{\vphantom L}_{\rm RFT}}[\,g\,] =
    -\frac\gamma{192\pi^2} \int d^4x\sqrt{g}\,R^2+O(\Re^3),
    \end{equation}
because all the terms depending on the parameter $\beta$ completely cancel out, and what remains coincides with the last quadratic term of \eqref{Gamma_quad}. This coincidence fully matches with the linear in curvature part of the trace anomaly \eqref{anomaly} (its $\gamma$-term) generated by the quadratic action \eqref{Gamma_quad}. Indeed, the conformal transformation of its nonlocal Weyl term contributes only to $O(\Re^2)$-part of the anomaly due to the fact that only its form factor $\ln(-\Box/\mu^2)$ is not Weyl invariant, and the whole $\gamma$-term of the anomaly entirely comes from the $R^2$-part of \eqref{Gamma_quad}.

\subsection{Fradkin-Vilkovisky gauge}

Another conformal gauge arises in context of conformal off-shell extension of Einstein gravity suggested in \cite{FV} and corresponds to the 4-dimensional version of the Yamabe problem. The representative of the conformal group orbit is chosen to be the metric with a vanishing scalar curvature
    \begin{equation} \label{FV_gauge}
    \chi_{{\vphantom L}_{\rm FV}}[\,\bar g\,] = \bar R,
    \end{equation}
which implies a nonlinear but still explicitly solvable equation for $\varSigma_{{\vphantom L}_{\rm FV}}$ ,
    \begin{equation}
    R[e^{-2\varSigma_{{\vphantom L}_{\rm FV}}}g_{\mu\nu}] = e^{3\varSigma_{{\vphantom L}_{\rm FV}}}\big(R-6\,\Box \big)\,e^{-\varSigma_{{\vphantom L}_{\rm FV}}} = 0.
    \end{equation}
This solution reads
    \begin{align}
    &\varSigma_{{\vphantom L}_{\rm FV}}=-\ln\Big(1+\frac16\frac1{\Box-R/6}R\Big), \label{varSigmaFV} \\
    &\lim\limits_{|x|\to\infty} e^{-\varSigma_{{\vphantom L}_{\rm FV}}} = 1
    \end{align}
in terms of the inverse of the conformal second order operator $\Box-\tfrac16 R$ subject to zero boundary conditions at infinity. This inverse operator also admits covariant curvature expansion and in the lowest order yields the function $\varSigma_{{\vphantom L}_{\rm FV}}$ coinciding with that of the RFT gauge \eqref{varSigma0},
    \begin{equation}
    \varSigma_{{\vphantom L}_{\rm FV}}=
    \varSigma_{{\vphantom L}_{\rm RFT}} + O(\Re^2),
    \end{equation}
and, therefore, generates in the quadratic order the same expression for the anomaly action
    \begin{equation}
    \varGamma_{{\vphantom L}_{\rm FV}}=
    \varGamma_{{\vphantom L}_{\rm RFT}} + O(\Re^3).
    \end{equation}

Using Eqs. \eqref{Gamma1-Gamma2} and \eqref{varSigmaRFT} it is easy to see that the difference between RFT and FV actions is given by the exact expression
    \begin{align}
    \varGamma_{{\vphantom L}_{\rm RFT}} - \varGamma_{{\vphantom L}_{\rm FV}} &=
    \frac1{16\pi^2}\int d^4x\,\sqrt{g}\big(\varSigma_{{\vphantom L}_{\rm RFT}}-\varSigma_{{\vphantom L}_{\rm FV}}\big) \nonumber \\
    &\times\Big[\,\alpha\, C^2+2\beta\,\Delta_4
    \big(\varSigma_{{\vphantom L}_{\rm RFT}}-\varSigma_{{\vphantom L}_{\rm FV}}\big)\Big],
    \end{align}
bilinear in the local Weyl squared term and conformally invariant nonlocal functional
    \begin{align}
    \varSigma_{{\vphantom L}_{\rm RFT}} - \varSigma_{{\vphantom L}_{\rm FV}}
    &= \frac14\,\frac1\Delta_4\calE_4 \nonumber \\
    &+ \ln\Big(1+\frac16\frac1{\Box-R/6}R\Big) = O(\Re^2). \label{SigmaRFT-SigmaFV}
    \end{align}
Therefore within perturbation theory these two actions remain coinciding even in the cubic order and become different only starting from the fourth order in the curvature.

Perturbatively both terms of \eqref{SigmaRFT-SigmaFV} produce similar nonlocal structures of tree-like nature, that is the terms characteristic of the tree-level approximation in field theory. Such terms are composed of the powers of inverse d'Alembertians acting on the curvature tensor structures or on the products of similar nonlocal tensor structures built according to the same pattern. However, taken separately as exact entities they have essentially different types of nonlocality. RFT action formalism involves the Green's function of the fourth order Paneitz operator, whereas the FV version of the action is based on the Green's function of the second order operator $\Box-\tfrac16 R$. Both operators are conformally covariant, but the Weyl transformation of $\Box-\tfrac16 R$ is different from \eqref{conf_Paneitz}
    \begin{equation}
    \Box-\tfrac16 R=
    e^{-3\sigma}\,\big(\bar\Box-\tfrac16\bar R\big)\,e^{\sigma},\qquad    g_{\mu\nu}=e^{2\sigma}\bar g_{\mu\nu}.
    \end{equation}
Moreover, FV action formalism involves a special logarithmic nonlinearity absent in RFT gauge fixing. The action of the Paneitz operator derivatives in \eqref{chi_action} can destroy this logarithmic structure, but the $\varSigma_{{\vphantom L}_{\rm FV}}C^2$-term in $\varGamma_{{\vphantom L}_{\rm FV}}$ still contains it intact.

A further comparison of the RFT and FV actions can be done along the lines of their ``naturalness''. RFT gauge \eqref{RFT_gauge} is based on structures organically belonging to the conformal anomaly formalism in the sense that it involves the same fundamental objects---the Branson curvature $\calE_4$ and the relevant Paneitz operator $\Delta_4$ which are immanently present in the flow of the anomalous action along the conformal group orbit \eqref{RTF}. One could even interpret this gauge as the one providing the extremum of $\beta$-terms in this expression with respect to the variation of the orbit parameter $\sigma$. This interpretation is, however, erroneous because $g_{\mu\nu}$, $\bar g_{\mu\nu}$ and $\sigma$ cannot be treated as independent variables in Eq. \eqref{RTF}.

On the contrary, FV gauge \eqref{FV_gauge} uses a somewhat extraneous entity---the scalar curvature---which is singled out only by the fact that it turns out to be the bearer of the metric conformal mode. As the result the advantage of FV gauge is that it does not involve higher than second order derivatives and does not produce double pole nonlocalities. Another advantage is that the equation \eqref{Gamma_vs_Gamma_chi} disentangling the FV anomaly action from the full effective action becomes in view of $\bar R=0$ much simpler
    \begin{equation} \label{Gamma_vs_Gamma_FV}
    \varGamma[\,g\,] = \varGamma_{{\vphantom L}_{\rm FV}}[\,g\,] + \varGamma[\,\bar g\,]\,\big|_{\bar g_{\mu\nu}[\,g\,]},
    \end{equation}
where $\bar g_{\mu\nu}[\,g\,] = e^{-2\varSigma_{{\vphantom L}_{\rm FV}}[\,g\,]}g_{\mu\nu}$, which is obviously consistent with the fact that $\varGamma_{{\vphantom L}_{\rm FV}}[\,\bar g\,]=0$ because $\varSigma_{{\vphantom L}_{\rm FV}}[\,\bar g\,]\equiv0$.

As compared to the FV version, among technical disadvantages of the RFT gauge and the action is the presence of fourth order derivatives of the Paneitz operator. Due to this the RFT version turns out to be vulnerable from the viewpoint of possible generalizations. For example, a modification of the gauge \eqref{RFT_gauge} by the additional Weyl squared term, $\chi_{{\vphantom L}_{\rm RFT}}\to\chi_{{\vphantom L}_{\rm RFT}}+aC^2$ would not work, because the relevant modification $\varSigma_{{\vphantom L}_{\rm RFT}}\to\varSigma_{{\vphantom L}_{\rm RFT}}+a\,(2\Delta_4)^{-1}C^2$ is badly defined for the reasons described above in the footnote \ref{footnote}---the additional term should have a total derivative structure.

The generalization to spacetimes with nontrivial topology is also not straightforward, because the condition \eqref{RFT_gauge} should not contradict nonvanishing Euler number of the manifold, which for compact manifolds {\em without} a boundary reads $e_E=\tfrac1{32\pi^2}\int d^4x\sqrt{g}\,E(x)$. Say, for a compact manifold of a finite volume $V=\int d^4x\,\sqrt{g}$ the gauge \eqref{RFT_gauge} can be chosen to be
    \begin{equation}
    \chi(\bar g) = \sqrt{\bar g}\,\Big(\bar E
    -\frac23\bar\Box \bar R
    -32\pi^2\frac{e_E}{\bar V}\Big), \label{RFT_gen_gauge}
    \end{equation}
but this leads to a nonlinear integro-differential equation for the relevant $\varSigma$
    \begin{align}
    4\sqrt{g}\Delta_4\varSigma &= \sqrt{g}\bigg(E-\frac23\Box R-32\pi^2 \frac{e^{-4\varSigma}}{\langle\,e^{-4\varSigma}\,
    \rangle}\frac{e_E}{V}\bigg), \\
    \langle\,e^{-4\varSigma}\,\rangle &\equiv
    \frac1{V}\int d^4x\,\sqrt{g} e^{-4\varSigma},   \label{Sigma_topology}
    \end{align}
which apparently can be solved analytically only by perturbations in $e_E/V$.

Unless stated otherwise, below we consider asymptotically flat spacetime with a trivial topology, whose Euler characteristics should be modified by the boundary term. For generic 4-dimensional manifolds with a smooth boundary it reads
    \begin{equation}
    e_E=\tfrac1{32\pi^2}\Big(\int_{\cal M} d^4x\sqrt{g}\,E(x)+\int_{\cal\partial M} d^3x\sqrt{\gamma}\varOmega(x)\Big), \label{Euler_surface}
    \end{equation}
where $\gamma=\det\gamma_{ab}$ and $\gamma_{ab}$ is the induced metric on $\partial{\cal M}$. For asymptotically flat case due to the contribution of $\partial{\cal M}$ at infinity $|\,x\,|\to\infty$ it equals $1$, so that everywhere in what follows the bulk part of the Euler characteristics is $\tfrac1{32\pi^2}\int d^4x\sqrt{g}\,E(x)\equiv e'_E=e_E-1=0$.\footnote{\label{Euler}I am grateful for this observation to M.Duff. Explicit and simple expression for the boundary term of the Euler characteristics in the 4-dimensional case can be found in \cite{Tseytlin_quadratic}, $\varOmega=\tfrac14R_{a\perp b\perp}K^{ab}+16\det K^a_b$, where $K_{ab}=\nabla_a n_b$ is the extrinsic curvature of the boundary, and $\perp$ denotes the projection on the outward pointing normal vector $n^\mu$. The last term in $\varOmega$ exactly reproduces the value of the Euler number $e_E=1$ for flat and asymptotically flat spaces  \cite{Christensen:1978tw}.}

\section{Conformal anomaly and covariant curvature expansion} \label{CurvExpansSect}

Despite the diversity of nonlocal structures of RFT and FV versions of anomaly action, neither of them seem to appear in conventional perturbation theory for quantum effective action. The covariant form of this perturbation theory in curved spacetime \eqref{Gamma_quad} was pioneered in \cite{DDI}, but its logarithmic nonlocal formfactor did not resemble the nonlocal operators of the RFT action \eqref{RFT}. Here we show how in spite of these discrepancies the anomaly action originates from covariant perturbation theory of \cite{CPTI, CPTII, CPTIII}.

This perturbation theory arose as a concrete implementation of the ideas of \cite{DDI} as an expansion in powers of covariant tensors of spacetime and fibre bundle curvatures and other covariant background field objects. This expansion is completely equivalent to standard Feynman diagrammatic technique and represents its resummation converting the original perturbation series in noncovariant odjects, like matter and metric field perturbations on top of flat and empty spacetime background, into the series in powers of covariant fields strengths denoted collectively below by $\Re$ and including spacetime and fibre bundle curvature.

To be more specific, consider the theory with the inverse propagator on top of the nontrivial field background $\hat F(\nabla)=F^A_B(\nabla)$, hat denoting the matrix structure of the operator acting in the space of fields $\varphi=\varphi^A(x)$ with a generic spin-tensor index $A$ and $\nabla=\nabla_\mu$ denoting the covariant derivative with respect to the corresponding fibre bundle connection,
\begin{equation}
\hat F(\nabla)=\Box+\hat P-\frac{\hat 1}6 R, \qquad \Box=g^{\mu\nu}\nabla_\mu\nabla_\nu.
\end{equation}
This operator is characterized by the ``curvatures''---metric Riemann tensor with its Ricci contractions, fibre bundle curvature $\hat{\cal R}_{\mu\nu}$ determining the commutator of covariant derivatives, $[\nabla_\mu,\nabla_\nu]\,\varphi
=\hat{\cal R}_{\mu\nu}\,\varphi$, and the potential term $\hat P$ (the term $-\tfrac{\hat 1}6 R$ is disentangled from the operator potential for reasons of convenience),
\begin{equation} \label{curvatures}
\Re=(R^\mu{}_{\nu\alpha\beta}, R_{\mu\nu}, R, \hat{\cal R}_{\mu\nu}, \hat P).
\end{equation}

In covariant perturbation theory the one-loop effective action gets expanded in powers of these curvatures
\begin{equation}
\varGamma = \frac12 {\rm Tr}\ln F(\nabla) = \!\!\!\! \stackrel{\text{local power div}}{\overbrace{\varGamma_0+\varGamma_1}}\!\!\!\!\!\!
+\,\varGamma_2+\varGamma_3 + O(\Re^4),
\end{equation}
where $\varGamma_n\sim\Re^n$. Within dimensional regularization of $2\omega$-dimensional spacetime, $\omega\to 2$, the zeroth and first order terms of the expansion represent pure power divergences (note that we consider the case of a massless theory, or the theory where the mass matrix is included in the potential term $\hat P$ and treated by perturbations), so that these two terms are annihilated by the regularization, while the second order term is given by the expression \cite{CPTI}
    \begin{align}
    \varGamma^{(2)}_{\rm dim\; reg} &= -\frac{\Gamma(2-\omega)\Gamma(\omega+1)\Gamma(\omega-1)}{2(4\pi)^\omega\Gamma(2\omega+2)}\,\mu^{4-2\omega} \nonumber \\
&\times\int dx\,\sqrt{g}\, {\rm tr}\,\Big\{R_{\mu\nu}(-\Box)^{\omega-2}R^{\mu\nu} \hat 1\nonumber\\
&-\frac1{18}(4-\omega)(\omega+1) R(-\Box)^{\omega-2}R \hat 1 \nonumber \\
&-\frac23(2-\omega)(2\omega+1)\,\hat P(-\Box)^{\omega-2}R \nonumber \\
&+2(4\omega^2-1)\,\hat P(-\Box)^{\omega-2}\hat P \nonumber\\
&+(2\omega+1) \hat{\cal R}_{\mu\nu}(-\Box)^{\omega-2}\hat{\cal R}^{\mu\nu}\Big\},  \label{Gamma_2}
    \end{align}
where $\omega=\tfrac{d}2\to 2$. Here ${\rm tr}$ denotes the matrix trace and the concrete coefficients implement the originally conjectured structure of dimensionally regularized effective action Lagrangian, $\Re(-\Box)^{\omega-2}\Re$, that was blueprinted in \cite{DDI}. What is important and should be especially emphasized is that $\Box=g^{\mu\nu}\nabla_\mu\nabla_\nu$ means here the full covariant d'Alembertian acting on a respective scalar $R$, tensor $R_{\mu\nu}$ or spintensor $\hat{\cal R}_{\mu\nu}$ and $\hat P$ objects.

For brevity we will consider the case of a single conformal scalar field with $\hat 1=1$, $\hat P=0$, $\hat{\cal R}_{\mu\nu}=0$ and the following values of the trace anomaly coefficients\footnote{The coefficients have the opposite sign to those of $b=-\alpha/16\pi^2$ and $b'=-\beta/16\pi^2$ in \cite{Duff:1977ay}, because in our case the stress tensor is defined with respect to the Euclidean effective action $\varGamma=-i\varGamma_L$ in contrast to the definition of $T^{\mu\nu}=2g^{-1/2}\delta\varGamma_L/\delta g_{\mu\nu}$ in the Lorentzian signature spacetime of \cite{Duff:1977ay}. Comparison with \cite{Birrell-Davies} should also take into account another sign of the stress tensor defined by the variation with respect to the contravariant metric.}
    \begin{equation}
    \alpha=-\frac1{120},\quad \beta=\frac1{360},
    \quad \gamma=-\frac1{180},          \label{single_scalar}
    \end{equation}
for which the action \eqref{Gamma_2} takes the form---a particular case of \eqref{Gamma_quad},
    \begin{align}
    \varGamma^{(2)}_{\rm ren} &= \frac1{32\pi^2} \int dx\,\sqrt{g}\,\Big\{\frac1{60}\Big[R_{\mu\nu}\gamma(-\Box)R^{\mu\nu} \nonumber \\
    &-\frac13R\gamma(-\Box)R\Big]+\frac{R^2}{1080}\Big\}\nonumber\\
    &=\frac1{32\pi^2} \int dx\,\sqrt{g}\,\Big\{\frac1{120}
    C_{\mu\nu\alpha\beta}\gamma(-\Box)C^{\mu\nu\alpha\beta} \nonumber \\
    &+\frac{R^2}{1080}\Big\}+O(\Re^3).          \label{Gamma_2_scal}
    \end{align}
Here $\gamma(-\Box)$ is the nonlocal formfactor (in minimal subtraction scheme with $\ln(4\pi)$ and Euler constants absorbed in $\mu$)
\begin{equation}
\gamma(-\Box)=\ln\Big(\!-\frac{\Box}{\mu^2}\Big)-\frac{16}{15}, \label{log}
\end{equation}
and the transition to the last line is valid up to the higher order terms in curvature and based on the nonlocal generalization of the identity
\begin{equation}
\int d^4x\,\sqrt{g}\,C^2=2\int d^4x\,\sqrt{g}\,(R_{\mu\nu}R^{\mu\nu}-\tfrac13 R^2)
\end{equation}
derived in \cite{CPTII,CPTIII} by integration by parts and use of the nonlocal representation of the Riemann tensor in terms of the Ricci one (see footnote \ref{Riemann} below).

The first term of this action is obviously conformal invariant in quadratic order, so that the linear in curvature part of the anomaly originates from the last term which is the RFT (or FV) action \eqref{RFT_quad} in the quadratic approximation with $\gamma=-1/180$. Thus, the RFT or FV action is fully recovered in this approximation from perturbation theory and, as expected, turns out to be local.

\subsection{Cubic order}

Quadratic order of the covariant curvature expansion is, in fact, a trivial generalization of the flat space expressions for self-energy operators of Feynman diagrammatic technique, because $\ln(-\Box/\mu^2)$ is just a straightforward replacement of the typical momentum space formfactor $\ln(p^2/\mu^2)$ by its position space version. At higher orders the situation becomes much more complicated and usually represented in terms of correlators of stress-tensor and other observables, written down in momentum space representation, see \cite{Skenderis2013,Skenderis2017,Skenderis2020} for the treatment of generic conformal field theories. These correlators are, of course, contained in the effective action expanded in curvatures which, for reasons of general covariance, we prefer to consider in coordinate representation.

In this representation the effective action becomes for each order $N$ in the curvature a sum of nonlocal monomials
    \begin{equation}
    \hspace{-0.3cm}\int d^4x_1\cdots d^4x_N\,F(x_1,\ldots,x_N)\nabla...\nabla\Re(x_1)...\Re(x_N)
    \end{equation}
with nonlocal multiple-point coefficients and covariant derivatives somehow acting on the product of curvatures at their various points. The absence of convenient and generally covariant momentum space representation makes us to work in coordinate representation and invent a special language which would simplify the formalism and make it manageable \cite{CPTI,CPTII,CPTIII}. This language is based on the operator representation of nonlocal formfactors,
    \begin{align}
    F(x_1,\ldots,x_N) &= \varGamma(\nabla_1,\ldots,\nabla_N) \nonumber \\
    &\times\delta(x_1,x_2)\,\delta(x_1,x_2)\cdots\delta(x_1,x_N),
    \end{align}
where $\varGamma(\nabla_1,\ldots,\nabla_N)$ is the operator valued function of $N$ independent covariant derivatives such that each $\nabla_i$ is acting on its own $x_i$. This allows one to write the orders of perturbation theory as
    \begin{align}
    \varGamma^{(N)} &= \frac1{2(4\pi)^2}\int\! d^4x\, \sqrt{g}\sum\limits_M \varGamma_M(\nabla_1,\ldots,\nabla_N) \nonumber \\
    &\times I_M(x_1,\ldots, x_N)\,\Big|_{\{x\}=x},
    \end{align}
where summation runs over all invariant monomials in curvatures of a given $n$-th order
    \begin{equation}
    I_M(x_1,\ldots,x_N) \sim \nabla\cdots\nabla\Re(x_1)\cdots\Re(x_N)
    \end{equation}
and after the action of all independent derivatives on their arguments all these arguments $\{x\}=(x_1,\ldots x_N)$ have to be identified.

In the cubic order for the full set of curvatures \eqref{curvatures} there are 29 such invariant structures built of these curvatures and their covariant derivatives with all indices fully contracted with each other. Moreover, in view of the scalar (no free indices) nature of the formfactors and the formal identity $\nabla_1+\nabla_2+\nabla_3=0$ (reflecting the possibility of integration by parts without surface terms, which is a counterpart to the momentum conservation in Feynman diagrams) the formfactors of $\varGamma^{(3)}$ can be written down as functions of three d'Alembertians $\Box_1$, $\Box_2$ and $\Box_3$ independently acting on three arguments of $I_M(x_1,x_2,x_3)$. Thus, cubic order reads as
    \begin{align}
    \varGamma^{(3)} &= \frac1{2(4\pi)^2}\int\! dx\, \sqrt{g} \sum\limits^{29}_{M=1}\varGamma_{M}(\Box_1,\Box_2,\Box_3) \nonumber \\
    &\times I_M(x_1,x_2,x_3)\,\Big|_{\{x\}=x}.
    \end{align}

The list of cubic invariants and their formfactors is presented in \cite{CPTIII,BMZ,MVZ}. It is very long and, as its details are not necessary for our purposes, we will not fully present it here. We only give the general structure of the nonlocal formfactors of these invariants. It reads as a sum of three different groups of terms
\begin{align}
    \varGamma_M(\Box_1,\Box_2,\Box_3) &= A_M\, \varGamma(\Box_1,\Box_2,\Box_3) \nonumber \\
   &\hspace{-0.5cm}+\sum_{1\leq i<k}^3\frac{D^{ik}_M} {(\Box_i-\Box_k)}\ln\frac{\Box_i}{\Box_k} + B_M.     \label{ff3}
\end{align}
Here $\varGamma(\Box_1,\Box_2,\Box_3)$ is the fundamental cubic formfactor corresponding to the triangular Feynman graph of massless theory with unit vertices \cite{CPTIIIa},
\begin{align}
&\varGamma(\Box_1,\Box_2,\Box_3) \nonumber \\
&= \int\limits_{\alpha\geq 0}
\frac{d^3\alpha\,\delta(1-\alpha_1-\alpha_2-\alpha_3)}{
\alpha_1\alpha_2(-\Box_3) + \alpha_1\alpha_3(-\Box_2) + \alpha_2\alpha_3(-\Box_1)},
\end{align}
which cannot be reduced to an elementary function. The operator-valued coefficients $A_M$, $B_M$ and $D_M^{ik}$ are  rational functions of three $\Box$-arguments with a polynomial numerator $P(\Box_1,\Box_2,\Box_3)$ and the denominator containing together with the product $\Box_1\Box_2\Box_3$ also the powers of a special quadratic form of these arguments $D$,
\begin{align}
&\hspace{-0.2cm}A_M,\; D^{ik}_M,\; B_M \sim \frac{P(\Box_1,\Box_2,\Box_3)}
{\Box_1\Box_2\Box_3\,D^L},\; L\leq 6,    \label{ADB}\\
&\hspace{-0.2cm}D = {\Box_1}^2\!+{\Box_2}^2\!+{\Box_3}^2\! -\! 2\Box_1\Box_2\! -\! 2\Box_1\Box_3\! -\! 2\Box_2\Box_3.
\end{align}

In this cubic order of the curvature expansion the conformal anomaly \eqref{anomaly}, which is quadratic in curvatures, was explicitly derived by the direct variation of the metric in \cite{anomaly_derivation}. Though this derivation has demonstrated nontrivial localization of the nonlocal terms under straightforward tracing the metric variational derivative, it still remained rather technical and not very illuminating because it has not revealed the anomalous part of the action. It turns out, however, that the transition to another basis of curvature invariants, suggested in \cite{MVZ,BMZ}, explicitly disentangles this part.

\subsection{Conformal resummation: Fradkin--Vilkovisky anomaly action}

The recovery of the anomaly part of the action and its conformal invariant part is based on a simple idea that the latter should consist of the series of Weyl invariant structures. The construction of Weyl invariants can be done by the gauge fixing procedure of the above type---choosing the representative metric on the group orbit by imposing the conformal gauge. Obviously the set of invariants surviving after imposing this gauge will be minimal if the gauge would explicitly annihilate the maximum number of invariants in their original full set. For this reason the FV gauge \eqref{FV_gauge} is much easier to use for the separation of the total set of invariants into the Weyl type ones and those which vanish when the gauge is enforced. As $R$ is one of the curvatures in the set of $\Re$ the FV gauge is more useful for the purpose of such a separation than the RFT gauge \eqref{RFT_gauge} which nonlinearly intertwines all the curvatures. Intuitively it is also clear because $R$, in contrast to $C^\alpha{}_{\beta\mu\nu}$, is a bearer of the conformal mode.

In the purely metric sector such a separation is attained by the transition to the new curvature basis \cite{MVZ},
\begin{equation}
\Re = \big(\,R^\mu_{\;\nu\alpha\beta},R_{\mu\nu},R\,\big) \to \tilde\Re = \big(\,C^\alpha_{\;\;\beta\mu\nu},R\,\big),
 \end{equation}
via expressing Ricci tensor in terms of the Weyl tensor and the Ricci scalar\footnote{\label{Riemann}In fact, the original basis and the curvature expansion of \cite{CPTI,CPTII,CPTIII} consisted of $R_{\mu\nu}$ and $R$ because in asymptotically flat Euclidean spacetime Riemann tensor can be expressed as nonlocal power series in the Ricci tensor,
$R_{\alpha\beta\mu\nu}=\frac1{\Box}\big(\nabla_\mu \nabla_\alpha R_{\nu\beta}-\nabla_\nu \nabla_\alpha R_{\mu\beta}\big) -(\alpha\leftrightarrow\beta)+O(\Re^2)$,---the corollary of contracted Bianchi identity.}. This expression follows from the contracted Bianchi identity which for the Weyl tensor reads as
\begin{equation}
\nabla^\beta\nabla^\alpha C_{\alpha\mu\beta\nu}=\frac12\Box R_{\mu\nu}-\frac16\nabla_\mu\nabla_\nu R
-\frac{g_{\mu\nu}}{12}\Box R + O(\Re^2).
 \end{equation}
This equation can be solved by iterations for Ricci tensor in terms of nonlocal series in powers of two objects---Ricci scalar $R$ and the new traceless (and up to quadratic order transverse) tensor $C_{\mu\nu}$ which is itself a nonlocal derivative of Weyl,
\begin{equation} \label{C}
C_{\mu\nu} = \frac2\Box\nabla^\beta\nabla_\alpha C^\alpha_{\;\;\mu\beta\nu}.
 \end{equation}
The resulting series begins with
 \begin{equation} \label{Ricci_vs_Weyl}
R_{\mu\nu} = C_{\mu\nu} + \frac13\nabla_\mu\nabla_\nu\frac1\Box R + \frac16 g_{\mu\nu}R + O(\tilde\Re^2).
 \end{equation}

Effective action reexpansion imples the transition from $I_M(x_1,\ldots, x_n)$ to a new basis of invariants
\begin{equation}
\tilde I_M(x_1,...x_n) \sim \nabla...\nabla\tilde\Re(x_1)...\tilde\Re(x_n),
\end{equation}
which can be separated in the set of monomials $I_C(x_1,\ldots, x_n)$ involving only $C_{\mu\nu}$ and the set of monomials $I_R(x_1,\ldots, x_n)$ containing at least one scalar curvature factor,
\begin{align}
I_C(x_1,...x_n) \sim &\nabla...\nabla C(x_1)... C(x_n),\\
I_R(x_1,...x_n) \sim &\nabla...\nabla R(x_1)C(x_2)... C(x_n), \nonumber \\
&\hspace{-0.5cm}\nabla...\nabla R(x_1)R(x_2)C(x_3)... C(x_n),\; ...
\end{align}

Expansion in the new basis of invariants implies, of course, the transition to a new set of their relevant formfactors
\begin{equation}
\varGamma_M(\nabla_1,...\nabla_n)\to\varGamma_C(\nabla_1,...\nabla_n), \varGamma_R(\nabla_1,...\nabla_n),
\end{equation}
and the new expansion takes the form
\begin{equation}
\varGamma=W+\varGamma_R,
\end{equation}
where $W$ is the Weyl and $\varGamma_R$ is the mixed Weyl--Ricci scalar parts of the whole expansion, which we write in abbreviated form (omitting multiple spacetime arguments and the operation of equating them)
    \begin{align}
    W &= \frac1{32\pi^2}\int\! d^4x\,\sqrt{g} \sum\limits_{n,C}\varGamma_C^{(n)}I_C^{(n)},\\
    \varGamma_R &= \frac1{32\pi^2}\int\! d^4x\,\sqrt{g} \sum\limits_{n,R}\varGamma_R^{(n)}I_R^{(n)}.
    \end{align}
Note that $W$ and its Weyl basis invariants are not Weyl invariant, because apart from Weyl tensors they contain covariant derivatives and nontrivial formfactors which do not possess conformal invariance properties.

The main statement on the conformal decomposition of the effective action of \cite{MVZ} is that
    \begin{equation} \label{conformization}
    \varGamma[\,g\,] = \varGamma_{{\vphantom L}_{\rm FV}}[\,g\,]
    + W[\,\bar g\,]\,\Big|_{\,\bar g_{\mu\nu} = e^{-2\varSigma_{{\vphantom L}_{\rm FV}}[\,g\,]} g_{\mu\nu}},
    \end{equation}
where $\varGamma_{{\vphantom L}_{\rm FV}}[\,g\,]$ is exactly the FV anomaly action introduced above\footnote{One can check that the last four lines of Eq. (24) in \cite{MVZ} form exact expression for $\varGamma_{{\vphantom L}_{\rm FV}}[\,g\,]$ by taking into account that the function $Z$ in this equation coincides with $-\varSigma_{{\vphantom L}_{\rm FV}}$ and satisfies the equation $\Box Z + \tfrac{1}{2}(\nabla Z)^2 = \tfrac{1}{3} R$.}. Conformally invariant part is obtained by the ``conformization'' of $W$, while the rest of the effective action is exhausted by the Fradkin-Vilkovisky anomaly action.  Invariant meaning of this representation is that the Ricci part of the full action is not independent, but fully determined by the anomaly and Weyl parts of the action. This representation looks as the realization of Eq.~\eqref{Gamma_vs_Gamma_FV} within perturbation theory in curvatures. This result is likely to resolve a long-standing debate between the proponents of the Riegert action and adherents of the flat space perturbation expansion for the effective action with typical nonlocal logarithmic form factors of the form \eqref{log}. Note that these form factors do not contribute to the anomaly even though their coefficients are directly related to its expression \eqref{anomaly}. Rather they become Weyl invariant under the substitution of $\bar g_{\mu\nu}$ as their functional argument.

Validity of the representation \eqref{conformization} was checked in the cubic order approximation for the effective action in \cite{MVZ}. The transition to the new basis of invariants in the second order leads to (see the second line of Eq.~\eqref{Gamma_2_scal}),
    \begin{align}
    \hspace{-0.5cm}W^{(2)}[\,g\,] &= \frac1{32\pi^2} \int dx\,\sqrt{g}\frac1{120} C_{\mu\nu\alpha\beta}\, \gamma(-\Box)C^{\mu\nu\alpha\beta},\\
    \varGamma^{(2)}_R[\,g\,] &= \frac1{32\pi^2} \int dx\,\sqrt{g}\frac1{1080}R^2,
    \end{align}
whereas in the third order it results in a great simplification of the ``Ricci scalar'' formfactors $\varGamma_R^{(3)}$ as compared to the original ones---they become much simpler and, moreover, in their expressions of the form \eqref{ff3} the coefficients $A,D^{ik}_M,B_M$ of \eqref{ADB} completely loose powers of the function $D$ in the denominator. Thus, modulo the contributions of $\ln(\Box_i/\Box_k)/(\Box_i-\Box_k)$ the formfactors $\varGamma_R^{(3)}$ acquire the tree-level structure. The terms with these factors get, however, completely absorbed with accuracy $O(\Re^4)$ by the replacement $W^{(2)}[\,g_{\mu\nu}\,]\to W^{(2)}[\,\bar g_{\mu\nu}\,]$ in view of the following relation \cite{CPTIII}
\begin{align}
&W^{(2)}[\,g\,] - W^{(2)}[\,\bar g\,] \nonumber \\
&\sim \int dx\, \sqrt{g}\, C_{\mu\nu\alpha\beta}\Big[\ln(-\Box)-\ln(-\bar\Box)\Big]C^{\mu\nu\alpha\beta} \nonumber \\
&= \int dx\, \sqrt{g} \frac{\ln(\Box_1/\Box_2)}{\Box_1-\Box_2}
[\Box_2-\bar\Box_2] C_{1\,\mu\nu\alpha\beta} C^{\mu\nu\alpha\beta}_2 + O(\Re^4), \nonumber \\
&\Box_2-\bar\Box_2\sim \Re_3+ O(\Re^2),
\end{align}
where the right hand side is the set of relevant cubic order terms with the above factor acting on two Weyl tensors out of three curvatures in $RCC$-type invariants. What remains in the sector of cubic $I^{(3)}_R$-invariants is the set of tree-like nonlocal form factors which comprise the curvature expansion of FV action up to $\tilde\Re^3$ order inclusive. This observation done in \cite{MVZ} can be formalized as the following sequence of identical transformations
    \begin{align}
    &\varGamma[\,g\,] = W^{(2+3)}[\,g\,] + \varGamma^{(2+3)}_R[\,g\,] + O(\Re^4)= W^{(2+3)}[\,\bar g\,] \nonumber \\
    &\;\;\;\; + \mathop{\underbrace{\varGamma^{(2+3)}_R[\,g\,] +\! \big(W^{(2)}[\,g\,] - W^{(2)}[\,\bar g\,] \big)}}_{{\varGamma^{(2+3)}_{{\vphantom L}_{\rm FV}} + O(\Re^4)}} + O(\Re^4),
    \end{align}
where the group of the last three terms forms Fradkin--Vilkovisky anomaly action expanded with $\tilde\Re^3$-accuracy. Explicitly the cubic part of $\varGamma_{{\vphantom L}_{\rm FV}}$ for the model of a single conformal scalar field with (\ref{single_scalar}) reads \cite{MVZ}
\begin{align}
\varGamma^{(3)}_{{\vphantom L}_{\rm FV}} &= -\frac1{32\pi^2} \int dx\,\sqrt{g}\bigg\{ \frac1{19440}\Big(\frac{2}{\Box}_3 - \frac{\Box_1}{\Box_2 \Box_3}\Big) R_1 R_2 R_3 \nonumber \\
&+ \frac{1}{1620\,\Box_2\Box_3} C_1^{\alpha\beta} \nabla_\alpha R_2 \nabla_\beta R_3 \nonumber \\
&+\frac1{540} \Big(\frac{4}{\Box}_2 - \frac{1}{\Box}_3 - \frac{2\,\Box_1}{\Box_2\Box_3} - \frac{\Box_3}{\Box_1\Box_2} \Big) C_1^{\mu\nu} C_{2\,\mu\nu} R_3 \nonumber \\
&+\frac1{135}\Big(\frac{1}{\Box_1\Box_2} - \frac{2}{\Box_2\Box_3} \Big) \nabla^\mu C_1^{\nu\alpha} \nabla_\nu C_{2\,\mu\alpha}R_3\nonumber\\
&- \frac{1}{135\,\Box_1\,\Box_2\Box_3} \nabla_\alpha\nabla_\beta C_1^{\mu\nu}\nabla_\mu\nabla_\nu C_2^{\alpha\beta} R_3\bigg\}\bigg|_{\,\{x\}=x},       \label{Gamma3}
\end{align}
where $C_{\mu\nu}$ is the ``Weyl'' part \eqref{C} of Ricci tensor \eqref{Ricci_vs_Weyl}

\subsection{The problem of double poles and global conformal transformations}

The expression \eqref{Gamma3} shows that in the cubic order the anomalous effective action is free from double pole nonlocal terms. For the FV action this is obviously true to all orders of the curvature expansion, since all its tree type nonlocalities originate from the Green's function of the conformal scalar operator $\Box-\tfrac16 R$. However, for the RFT action double poles formally appear starting from the fourth order in the curvature because the metric variation of $\varSigma_\chi=\varSigma_{\rm RTF}$ in \eqref{chi_action} leads to the action of the inverse Paneitz operator upon the square of the Weyl tensor $C^2 = C_{\mu\nu\alpha\beta}C^{\mu\nu\alpha\beta}$ due to a formal variational rule
\begin{equation}
\hspace{-0.2cm}\int d^4x\,\sqrt{g}\,C^2\delta\varSigma_{\rm RFT} = \int d^4x\,\sqrt{g}\,(\Delta_4^{-1}C^2)\delta(\ldots).
\end{equation}
This operation is not well defined, because $C^2$ is not a total derivative and the repeated action of $1/\Box$ upon generic test functions in four dimensions leads to IR divergent integrals---see footnote \ref{footnote}. In the cubic order of $\varGamma_{\rm RFT}$ this problem does not arise because of the extra $\Box$ factor in $\Box R$, as it was checked in \cite{Coriano-Maglio-Mottola} by explicit calculations of $\langle\,TTT\,\rangle$ correlators, but one is not granted to be free from this difficulty for higher order correlators.

In fact this is a typical situation of IR divergences in two dimensions, where the kernel of $1/\Box$ has a logarithmic dependence at infinity, and the correlators of undifferentiated conformal fields $\phi$ are UV divergent, while the correlators $\langle\partial\phi(x)\partial\phi(y)\cdots\rangle$ stay well defined. Apparently, the same property in four dimensions also underlies absence of unitarity in dipole theories with $1/\Box^2$-type propagators recently discussed in \cite{Tseytlin}. The mechanism of transition from operators to their derivatives in shift symmetric theories actually helps to justify the RFT action as a source of well defined stress tensor correlators and extend the validity of results in \cite{Coriano-Maglio-Mottola} to all higher orders.

This follows from the observation that the Paneitz operator reads
    \begin{equation}
\sqrt{g}\Delta_4 = \partial_\mu \Big[\sqrt{g}\big(\nabla^\mu\nabla_\nu + 2R^{\mu\nu} - \frac{2}{3}Rg^{\mu\nu}\big)\Big]\partial_\nu
    \end{equation}
and, therefore, perturbatively on the flat space background can be represented as
    \begin{equation}
    \sqrt{g}\Delta_4 = \tilde\Box^2 + V, \; \tilde\Box = \delta^{\mu\nu}\partial_\mu\partial_\nu,\; V = \overrightarrow{\partial_\mu V^{\mu\nu}\partial_\nu},
    \end{equation}
where the perturbation $V = O(\Re)$ has a special form---another differential operator $V^{\mu\nu}$ sandwiched between two derivatives with all derivatives acting to the right (which is indicated by the arrow). Within perturbation theory in powers of $V$ the action of the inverse operator on a generic test function $\psi$---scalar density---could have been understood as the expansion
    \begin{align}
    \phi = \frac1{\sqrt{g}\Delta_4}\psi &= \sum\limits_{n=0}^\infty\frac{(-1)^n}{\tilde\Box^2} \left(V\frac1{\tilde\Box^2}\right)^n\psi \nonumber \\
&= \sum\limits_{n=0}^\infty\frac{(-1)^n}{\tilde\Box^2} \left(\overrightarrow{\partial_\mu V^{\mu\nu}} \frac1{\tilde\Box^2}\partial_\nu\right)^n\psi,
    \end{align}
where we deliberately permuted the factors of $\partial_\nu$ and $1/\tilde\Box^2$ using their formal commutativity in order to provide the action of $1/\tilde\Box^2$ on the total derivative function. Thus all terms of this expansion except the first one become infrared finite. The first term $(1/\tilde\Box^2)\psi$, however, makes this function $\phi$ ill defined. On the contrary, its derivative $\partial_\alpha\phi$ becomes consistent if one understands the first term of the expansion as $(1/\tilde\Box^2)\partial_\alpha\psi$, so that the prescription for the operation of $\partial_\alpha(1/\sqrt{g}\Delta_4)$ on a generic non-derivative type test function reads as
    \begin{equation} \label{prescription}
    \partial_\alpha\frac1{\sqrt{g}\Delta_4}\psi=
    \sum\limits_{n=0}^\infty\frac{(-1)^n}{\tilde\Box^2}\partial_\alpha
    \left(\overrightarrow{\partial_\mu V^{\mu\nu}}
    \frac1{\tilde\Box^2}\partial_\nu\right)^n\psi.
    \end{equation}

With this prescription the term $C^2\varSigma_{\rm RFT}$ in the RFT action becomes perturbatively well defined to all orders of expansion. Indeed, this term with $\varSigma_{\rm RFT}$ given by \eqref{varSigmaRFT} and on account of total derivative structure $\sqrt{g}(E-\tfrac23\,\Box R\big)=\partial_\alpha E^\alpha$ can be rewritten by integration by parts as
    \begin{equation}
    \hspace{-0.2cm}4\!\int d^4x\sqrt{g}\,C^2\varSigma_{\rm RFT}=-\int d^4x\,\sqrt{g}E^\alpha\,\partial_\alpha\frac1{\sqrt{g}
    \Delta_4}\big(\sqrt{g}\,C^2\big)
    \end{equation}
with the above prescription \eqref{prescription}. This confirms a well defined nature of all multiple point correlators of stress tensor generated by RFT action.

Finally, it is worth discussing the effective action behavior under global conformal transformations with $\sigma_0={\rm const}$. Higher order curvature terms of the effective action scale as negative powers of $e^{\sigma_0}$ and therefore are irrelevant in the IR limit. In \cite{Mazur-Mottola2001} this was a main argument in favor of a dominant role of the Wess--Zumino action \eqref{RTF} in this limit because $\Delta\varGamma[g, \sigma]$ behaves linearly in $\sigma_0$ (or logarithmically in the distance). Indeed,
    \begin{align}
    \Delta\varGamma[g, \sigma + \sigma_0] &= \Delta\varGamma[g, \sigma] \nonumber \\
    &+\sigma_0\Big(\frac\gamma{32\pi^2}\int d^4x\,\sqrt{g}\,C^2 + \beta\,e'_E\Big),  \label{global}
    \end{align}
where $e'_E$ is the Euler characteristics of the manifold modulo its boundary contribution (see footnote \ref{Euler}). Note, however, that this behavior cannot be captured within the nonlocal RTF form of the anomaly action \eqref{RFT} because it is valid only under Dirichlet boundary conditions for the Green's function of $\Delta_4$ (which would be violated by the $\sigma_0$-shift). In other words, the expression \eqref{RFT} lacks the contribution of the zero mode of the Paneitz operator, which on the contrary is explicitly featuring in \eqref{global}. For compact manifolds with possibly nontrivial topology global Weyl transformations would not contradict boundary conditions, and these transformations will obviously show up in the generalized RFT gauge \eqref{RFT_gen_gauge} as an ambiguity of the solution for Eq.~\eqref{Sigma_topology}, $\varSigma\to\varSigma+\sigma_0$.

\section{Stress tensor in conformally related spacetimes} \label{StressTensorSect}

Equations \eqref{Gamma_vs_Gamma_chi} and \eqref{Gamma_vs_Gamma_FV} show that the anomalous action makes sense as an object specifying the difference of effective actions on conformally related metrics and other fields. Outside of this context this action, being a subject of shifting by an arbitrary conformal invariant functional $W^{\rm conf}[\,g\,]$, as in Eq.~\eqref{shifting}, is not very instructive because such a shift can include essential physical information on conformally invariant degrees of freedom. Anomaly action $\varGamma_\chi$, or it would be better to say, the Wess--Zumino type action \eqref{RTF}---the generating functional of $\varGamma_\chi$---is really useful in situations when the physics of a conformally related spacetime with the metric $\bar g_{\mu\nu}$ is fully known. Then the effective action at $g_{\mu\nu}$ can be completely recovered from the knowledge of the Weyl anomaly.

The simplest situation belongs to the class of conformally flat spacetimes when $\bar g_{\mu\nu}$ can be associated with flat metric for which all the metric field invariants are vanishing and $\varGamma[\,\bar g\,]$ is either exactly zero or calculable for quantum matter fields in flat spacetime. In particular, the fundamental observable which can then be obtained is the UV renormalized expectation value of the stress tensor of classically conformally invariant fields,
    \begin{equation}
    \sqrt{g}\,\big\langle\,
    T^{\alpha\beta} \big\rangle=2
    \frac{\delta\varGamma_{\rm ren}}{\delta g_{\alpha\beta}}
                     \label{stress}
    \end{equation}
provided $\langle\,\bar T^{\alpha\beta}\rangle=0$ or known from flat space physics. Here we derive from \eqref{RTF} the expression for the difference of (densitized) stress tensors $\sqrt{g}\,\langle\,T^\alpha_\beta\rangle - \sqrt{\bar g}\,\langle\,\bar T^\alpha_\beta\rangle$, which for a conformally flat spacetime coincides with a well-known Brown--Cassidy expression \cite{Brown-Cassidy} and generalizes it to the case of a nonvanishing Weyl tensor.

\subsection{Conformal anomaly from the divergent part of the effective action}

To derive the behavior of the renormalized stress tensor on the conformal group orbit we, first, have to trace the origin of conformal anomaly as the result of subtracting UV divergences from covariantly regularized effective action, $\varGamma_{\rm ren}=\varGamma_{\rm reg}-\varGamma_\infty$. In dimensional regularization, $\varGamma_{\rm reg}={}^{(d)}\!\varGamma$, these divergences are given by
    \begin{align}
    \varGamma_\infty &= -\frac{1}{16\pi^2\epsilon} \int d^d x\,\sqrt{g}\,a_2 \nonumber \\
    &= \frac{1}{16\pi^2\epsilon} \int d^d x\,\sqrt{g}\,\big(\alpha\, {}^{(4)}\!C^2+\beta\, {}^{(4)}\!E\,\big), \label{Gamma_div}
    \end{align}
where $\epsilon = 4-d$, ${}^{(4)}\!C^2$ and $^{(4)}\!E$ are the four-dimensional invariants formally continued to $d$-dimensions and $a_2$ is the relevant second Schwinger--DeWitt coefficient of the corresponding heat kernel expansion for the inverse propagator of the theory \cite{SchDW,PhysRep,Scholarpedia},
    \begin{align}
    &\;a_2 = -\big(\alpha\, {}^{(4)}\!C^2 +\beta\, {}^{(4)}E+\gamma\,\Box R\big), \label{a2}\\
    &{}^{(4)}C^2 = R_{\mu\nu\alpha\beta}^2 - 2R_{\mu\nu}^2 + \frac13 R^2,\\
    &{}^{(4)}\!E = R_{\mu\nu\alpha\beta}^2 - 4R_{\mu\nu}^2 + R^2.
    \end{align}
This structure of $a_2$ follows from the local conformal invariance of the pole residue of $\varGamma_\infty$ at $d=4$ and associated with the integrability (or conformal Wess--Zumino) condition for a conformal anomaly. It includes the topological Gauss--Bonnet density $\sqrt{g}E$, Weyl tensor squared and the total derivative $\Box R$ terms.

Conformal anomaly arises as a contribution of the conformal transformation of the one-loop counterterm \eqref{Gamma_div} subtracted from the regularized effective action
    \begin{equation}
    \sqrt{g}\,\big\langle\,
    T^\alpha_\alpha \big\rangle=-2g_{\alpha\beta}
    \frac{\delta\varGamma_\infty}
    {\delta g_{\alpha\beta}},                 \label{anom_from_div}
    \end{equation}
because the regularized (but not yet renormalized by counterterm subtracting) action $\varGamma_{\rm reg}$ is assumed to be conformally invariant\footnote{\label{reg_Weyl}Or the Weyl invariance violation of dimensionally regularized $\varGamma_{\rm reg}$ is proportional to $(d-4)^2$ as it happens for spin one case \cite{Duff:1977ay}, so that it does not contribute to the residue of the simple pole in dimensionality.}. The $\Box R$ term does not contribute to the divergences but it appears in the conformal anomaly in view of the conformal transformation of the Weyl squared term continued to $d$ dimensions. Moreover, within the above subtraction scheme its coefficient $\gamma$ in the anomaly turns out to be determined by the coefficient $\alpha$ of the Weyl term \cite{Duff:1977ay}.

Indeed, introduce conformally covariant Weyl tensor in $d$ dimensions
\begin{align}
{}^{(d)}\!C_{\mu\nu\alpha\beta} &= R_{\mu\nu\alpha\beta} +2P_{\beta[\mu}g_{\nu]\alpha} - 2P_{\alpha[\mu}g_{\nu]\beta}, \\
{}^{(d)} C^\mu{}_{\nu\alpha\beta} &= {}^{(d)}\!\bar C^\mu{}_{\nu\alpha\beta},      \label{Schouten}
\end{align}
which is written down in terms of the Schouten tensor
\begin{equation}
P_{\mu\nu} \equiv \frac{1}{d-2}\Big(R_{\mu\nu} - \frac{Rg_{\mu\nu}}{2(d-1)}\Big).
\end{equation}
In view of the relation between the square of Weyl tensors $^{(d)}C^2\equiv{}^{(d)}C_{\mu\nu\alpha\beta}^2$ and $C^2\equiv {}^{(4)}\!C^2_{\mu\nu\alpha\beta}$ (both formally continued to $d$ dimensions) \cite{Brown-Cassidy}
\begin{equation} \label{C^2_vs_Cd^2}
{}^{(4)}\!C^2 = {}^{(d)}\!C^2 - \frac{\epsilon}2 \big(E-C^2-\tfrac19 R^2\big) + O(\epsilon^2)
\end{equation}
one has
    \begin{align}
    &\frac\delta{\delta g_{\mu\nu}}\int d^dx\,\sqrt{g}\,C^2 = \frac\delta{\delta g_{\mu\nu}}\int d^dx\,\sqrt{g}\, {}^{(d)}\!C^2\nonumber\\
    &\;\;\;\;+\frac{\epsilon}2\frac\delta{\delta g_{\mu\nu}}
    \int d^4 x\,\sqrt{g}\,\big(C^2+\tfrac19 R^2\big) + O(\epsilon^2).                                  \label{var_C^2}
    \end{align}
Then, since the tensor $^{(d)}C_{\mu\nu\alpha\beta}$ is conformally covariant in any dimension, $g_{\mu\nu}(\delta/\delta g_{\mu\nu}) \int d^dx\,\sqrt{g}\, ^{(d)}C^2 = -\tfrac{\epsilon}2\sqrt{g}\, {}^{(d)}C^2$, we have
    \begin{equation}
    \frac1{\epsilon}g_{\mu\nu}\frac\delta{\delta g_{\mu\nu}}\int d^dx\,\sqrt{g}C^2 = -\frac12 \sqrt{g}\Big(C^2+\frac23\Box R\Big) +O(\epsilon).
    \end{equation}
Using this in \eqref{anom_from_div} one recovers the $C^2$ and the $\Box R$ terms in the expression for the anomaly
\begin{equation} \label{trace_anomaly}
\sqrt{g}\,\big\langle T^\alpha_\alpha \big\rangle = -\frac{1}{16\pi^2}\sqrt{g}\,a_2,
\end{equation}
with the parameter $\gamma$ related to the coefficient $\alpha$ of the Weyl squared term \cite{Duff:1977ay}
\begin{equation} \label{gamma_vs_alpha}
\gamma=\frac23\alpha.
\end{equation}
This simple expression for the trace anomaly in terms of the second Schwinger--DeWitt coefficient also follows from the zeta-function regularization \cite{DeWitt1980}.

The Gauss--Bonnet part of the anomaly follows from the conformal variation of the ${}^{(4)}\!E$-term in the divergent part of the action. Just like $\Box R$, as the residue of the pole in $\varGamma_\infty$ the integral of $\sqrt{g}{}^{(4)}\!E$ at least naively does not contribute to the stress tensor, because in four dimensions this integral is a constant Euler characteristics of the manifold. But in a covariant renormalization procedure the coefficient of $1/\epsilon$ in $\varGamma_\infty$ cannot be treated other than as a $d$-dimensional object, so that $\int d^dx\sqrt{g}{}^{(4)}\!E$ is no longer a topological invariant, and its metric variation is nontrivial. Therefore, rewriting, similarly to \eqref{C^2_vs_Cd^2}, the dimensionally continued Gauss--Bonnet density in terms of ${}^{(d)}\!C^2$,
\begin{align}
{}^{(4)}\!E &= R_{\mu\nu\alpha\beta}^2 - 4R_{\mu\nu}^2 + R^2 \nonumber \\
&= {}^{(d)}\!C^2 - (2-3\epsilon)\big(R_{\mu\nu}^2 - \tfrac13 R^2\big)+O(\epsilon^2),
\end{align}
one has
    \begin{align} \label{var_E}
    \frac1{\epsilon}\frac\delta{\delta g_{\alpha\beta}}\int d^dx\,\sqrt{g}\, {}^{(4)}\!E =& -\sqrt{g}\big(\,\tfrac12W^{\alpha\beta}+^{(3)}\!\!H^{\alpha\beta}\nonumber\\
    &+2R_{\mu\nu}C^{\mu\alpha\nu\beta}\big) + O(\epsilon),
    \end{align}
where the two new tensors arise
\begin{align}
&\!\!\!{}^{(3)}\!H^{\alpha\beta} =  R^{\alpha\mu}R^\beta_\mu - \frac{2}{3}RR^{\alpha\beta} - \frac{1}{2}g^{\alpha\beta}R_{\mu\nu}^2 + \frac{1}{4}g^{\alpha\beta}R^2,  \label{H3}\\
&W^{\alpha\beta} = \lim\limits_{\epsilon\to 0} \frac{1}{\epsilon} \left(4\,{}^{(d)}\!C^\alpha{}_{\mu\nu\lambda} {}^{(d)}\!C^{\beta\mu\nu\lambda} - g^{\alpha\beta}\,{}^{(d)}\!C^2 \right).  \label{W}
\end{align}
The limit to $d=4$ for the tensor $W^{\alpha\beta}$ is regular here because at $d=4$ there is the important identity
\begin{equation}
4\,{}^{(4)}\!C^\alpha{}_{\mu\nu\lambda} {}^{(4)}\!C^{\beta\mu\nu\lambda}= g^{\alpha\beta}{}^{(4)}\!C^2
\end{equation}
---it can be proven by antisymmetrization over five indices in the four-dimensional spacetime \cite{basis}. Tensors ${}^{(3)}H^{\alpha\beta}$ and $W^{\alpha\beta}$ have the following traces
\begin{equation} \label{trace_W}
{}^{(3)}\!H^\alpha_\alpha=\tfrac13R^2 -R_{\mu\nu}^2=\tfrac12(E-C^2), \quad W_\alpha^\alpha = C^2.
\end{equation}
Thus from \eqref{var_E} and \eqref{trace_W} we have the relation
    \begin{equation}
    \frac2{\epsilon}g_{\alpha\beta}\frac\delta{\delta g_{\alpha\beta}}\int d^dx\,\sqrt{g}\, {}^{(4)}\!E = -\sqrt{g}{}^{(4)}\!E + O(\epsilon),
    \end{equation}
which recovers the contribution of $E$-term in the conformal anomaly \eqref{trace_anomaly} with the expression \eqref{a2} for $a_2$.

\subsection{Minimal form of Wess-Zumino action and {\em a}-theorem}

Of course there is a big ambiguity in the above analytic continuation of the coefficients relating 4-dimensional objects to their $d$-dimensional counterparts. This ambiguity reduces to the renormalization by finite 4-dimensional counterterms $\int d^4x\sqrt{g}\,R_{\mu\nu\alpha\beta}^2$, $\int d^4x\sqrt{g}\,R_{\mu\nu}^2$ and $\int d^4x\sqrt{g}\,R^2$ among which in view of the total-derivative nature of the Gauss-Bonnet density only one counterterm can additionally break Weyl invariance and change the coefficient $\gamma$ of the $\Box R$ term in the conformal anomaly. This is because the combination $\int d^4x\,\sqrt{g}(C^2-E)=2\int d^4x\,\sqrt{g}(R_{\mu\nu}^2-\tfrac13 R^2)$ is Weyl invariant, and such a counterterm can be chosen as the square of the curvature scalar, satisfying
    \begin{equation}
    g_{\mu\nu}\frac\delta{\delta g_{\mu\nu}}\int d^4x\sqrt{g}\,R^2
    = -6\sqrt{g}\,\Box R.
    \end{equation}
Therefore this {\em finite local} counterterm can be used to alter the coefficient $\gamma$ and, in particular, put it to zero by a special finite renormalization which we will denote by a subscript $\rm Ren$,
    \begin{equation} \label{renormaction1}
    \hspace{-0.2cm}\varGamma_{\rm ren}[\,g\,]\to \varGamma_{\rm Ren}[\,g\,]
    \equiv\varGamma_{\rm ren}[\,g\,] + \frac\gamma{192\pi^2}\int d^4x
    \sqrt{g}\,R^2.
    \end{equation}

Regularization and subtraction scheme dependence of $\gamma$-coefficient manifests itself in the violation of the relation \eqref{gamma_vs_alpha} for the dimensionally regularized electromagnetic vector field \cite{Birrell-Davies}, but ultimately does not change the physics of the theory because of the locality of the covariant counterterm $\int d^4x\sqrt{g}\,R^2$, whose subtraction point should be determined from the comparison with the observable value of its coupling constant. In the cosmological example considered below the above renormalization \eqref{renormaction1} corresponds to fixing the coupling constant in the Starobinsky $R^2$-model \cite{slih_R^2}.

The renormalization (\ref{renormaction1}) has an important consequence -- with $\gamma=0$ the terms with quartic derivatives of $\sigma$, contained in the combination $\tfrac\beta{16\pi^2}\int d^4x\,\big(4\sqrt{\bar g}\,\sigma\bar\Delta_4\sigma-\frac19\sqrt{g}\,R^2\big)$ of (\ref{RTF}), completely cancel out, and the resulting {\em minimal} Wess-Zumino action  does not acquire extra hihger-derivative degrees of freedom,
    \begin{eqnarray}
    &&\varGamma_{\rm Ren}[\,g\,]-\varGamma_{\rm Ren}[\,\bar g\,]
    =\frac\alpha{16\pi^2}\int d^4x\,\sqrt{\bar g}\,
    \bar C_{\mu\nu\alpha\beta}^2\sigma             \nonumber\\
    &&\quad
    +\frac{\beta}{16\pi^2}\int d^4x\,\sqrt{\bar g}\,
    \Big\{\bar E\,\sigma-4\,\big(\bar R^{\mu\nu}
    -\tfrac12\bar g^{\mu\nu}\bar
    R\,\big)\,\partial_\mu\sigma\,\partial_\nu\sigma  \nonumber\\
    &&\quad-4\,\bar\Box\sigma\,
    (\bar\nabla^\mu\sigma\,\bar\nabla_\mu\sigma)
    -2\,(\bar\nabla^\mu\sigma\,
    \bar\nabla_\mu\sigma)^2\Big\}.         \label{minimal}
    \end{eqnarray}
    
This minimal version of the action for the dilaton field $\sigma$ was discussed in \cite{Jack-Osborn} and used in the derivation of the $a$-theorem in \cite{KomargodskiSchwimmer,Komargodski} -- monotonically decreasing coefficient $a=\beta/16\pi^2$ in the RG flow of the theory from UV to IR domains. This theorem is based on the sign of the last quartic interaction term for this field, related to the cross section of the forward $2\to 2$ dilaton scattering which should be positive in unitary theory, its unitarity being related to the absence of higher-derivative ghosts in (\ref{minimal}).

\subsection{Renormalized stress tensors}

The behavior of the stress tensor on the orbit of the conformal group can be obtained by using the commutativity of the following functional variations
\begin{equation}
\left[\,g_{\mu\nu}(y)\frac\delta{\delta g_{\mu\nu}(y)},
g_{\beta\gamma}(x)\frac\delta{\delta g_{\alpha\gamma}(x)}\right]=0,
\end{equation}
which allows one to write
\begin{align}
&\frac\delta{\delta\sigma(y)}\sqrt{g}\left\langle T^\alpha_\beta(x) \right\rangle = 2g_{\beta\gamma}(x)\frac\delta{\delta g_{\alpha\gamma}(x)}
\frac{\delta\varGamma_{\rm ren}}{\delta\sigma(y)}\bigg|_{g_{\mu\nu} = e^{2\sigma}\bar g_{\mu\nu}} \nonumber \\
&\qquad= g_{\beta\gamma}(x)\frac\delta{\delta g_{\alpha\gamma}(x)}
 \sqrt{g}(y)\left\langle T^\mu_\mu(y)\right\rangle
\Big|_{g_{\mu\nu} = e^{2\sigma}\bar g_{\mu\nu}}.
\end{align}
Bearing in mind that $g_{\beta\gamma}\delta/\delta g_{\alpha\gamma}=
\bar g_{\beta\gamma}\delta/\delta\bar g_{\alpha\gamma}$ at fixed $\sigma$ and functionally integrating this relation over $\sigma$ one has
\begin{equation}
\sqrt{g}\,\big\langle T^\alpha_\beta \big\rangle - \sqrt{\bar g}\,\big\langle\bar T^\alpha_\beta \big\rangle
= 2\bar g_{\beta\gamma} \frac\delta{\delta\bar g_{\alpha\gamma}}
\Delta\varGamma[\,\bar g,\sigma\,],    \label{var_WZ}
\end{equation}
where $\Delta\varGamma[\,\bar g,\sigma\,] = \varGamma_{\rm ren}-\bar\varGamma_{\rm ren}$ is given by \eqref{RTF}.

Before calculating this difference by the metric variation of $\Delta\varGamma[\bar g, \sigma]$ it is instructive to obtain it directly from the divergent part of the action as it was done in \cite{Brown-Cassidy}. Note that $\varGamma_{\rm ren}-\bar\varGamma_{\rm ren} = -(\varGamma_\infty-\bar\varGamma_\infty)$ because $\varGamma_{\rm reg}$ does not contribute to the anomaly (see footnote \ref{reg_Weyl}). Therefore,
\begin{equation}
\sqrt{g}\,\big\langle\,T^\alpha_\beta \big\rangle\,\Big|_{\,\bar g}^{\,g} = -2\,g_{\beta\gamma}\frac{\delta\varGamma_\infty}{\delta g_{\alpha\gamma}}\,\Big|_{\,\bar g}^{\,g}
\end{equation}

To calculate the contribution of the $^{(4)}C^2$-term in $\varGamma_\infty$ we rewrite it in terms of $^{(d)}C^2$ and use Eq.~\eqref{var_C^2}. This leads to the contribution of the first term of this equation
    \begin{align}
    &\frac\delta{\delta g_{\mu\nu}}\int d^dx\,\sqrt{g}\, {}^{(d)}C^2 = -\frac{\epsilon}2 \sqrt{g}\,W^{\mu\nu}-4\sqrt{g}\,{}^{(d)}B^{\mu\nu},\\
    &{}^{(d)}B^{\mu\nu}=\Big(\frac1{d-2}R_{\alpha\beta}
    +\nabla_{(\alpha}\nabla_{\beta)}\Big)C^{\mu\alpha\nu\beta},
    \end{align}
where the tensor $W^{\mu\nu}$ is defined by Eq.~\eqref{W} and ${}^{(d)}\!B^{\mu\nu}$ is the $d$-dimensional Bach tensor. Assembling this with the second term of Eq.~\eqref{var_C^2} we get on the orbit of the conformal group
    \begin{multline}
    \frac1{\epsilon}g_{\beta\gamma}\frac\delta{\delta g_{\alpha\gamma}}\int d^dx\,\sqrt{g}\, ^{(4)}\!C^2\,\Big|_{\,\bar g}^{\,g} \\
    = -\sqrt{g}\Big[\,\frac{4}{\epsilon} {}^{(d)}\!B^\alpha_\beta
    +\frac1{18}{}^{(1)}\!H^\alpha_\beta\,\Big]_{\,\bar g}^{\,g} + O(\epsilon),                              \label{var_C^2_0}
    \end{multline}
where the tensor ${}^{(1)}\!H^\alpha_\beta$ is given by the equation
    \begin{align}
    &{}^{(1)}\!H^\alpha_\beta = \frac1{\sqrt{g}}g^{\alpha\gamma}\frac\delta{\delta g^{\beta\gamma}}\int d^4x\,\sqrt{g}R^2 \nonumber \\
&= -\frac{1}{2}\delta^\alpha_\beta R^2 + 2RR^\alpha_\beta +2\delta^\alpha_\beta \Box R - 2\nabla^\alpha\nabla_\beta R,
    \end{align}
and we took into account that the both tensor densities $\sqrt{g}\,W^\alpha_\beta$ and $\sqrt{g}\,B^\alpha_\beta$ {\em in four dimensions} are invariant on the conformal orbit. Outside of four dimensions the Bach tensor density transforms on this orbit as (here as above $g_{\mu\nu}=e^{2\sigma}\bar g_{\mu\nu}$)
    \begin{equation}
    \sqrt{g}\,{}^{(d)}\!B^\alpha_\beta\,
    \Big|_{\bar g}^g = -\frac{\epsilon}2 \sqrt{\bar g}\big(\bar R^{\mu\nu}+2\bar\nabla^{(\mu}\bar\nabla^{\nu)}\big)\big(\sigma\bar C^\alpha{}_{\mu\beta\nu}\big)
    +O(\epsilon^2),         \label{new}
    \end{equation}
which makes the first term on the right hand side of \eqref{var_C^2_0} well defined at $d\to 4$. Note that the expression $\sqrt{\bar g}(\bar R^{\mu\nu}+2\bar\nabla^{(\mu}\bar\nabla^{\nu)}\big)\big(\sigma\bar C^\alpha{}_{\mu\beta\nu})$ treated as a functional of independent $\bar g_{\mu\nu}$ and $\sigma$ is Weyl invariant under local conformal transformations of the barred metric. This can be easily inferred from the invariance of Eq.(\ref{new}) under the interchange $g_{\mu\nu}\leftrightarrow\bar g_{\mu\nu}$ and $\sigma\to -\sigma$ or directly checking the conformal transformation of $\bar g_{\mu\nu}$ (with a fixed scalar $\sigma$).

The contribution of Gauss--Bonnet term to the stress tensor behavior on the conformal orbit is obtained from using \eqref{var_E}--\eqref{H3}. Collecting this contribution with the contribution (\ref{var_C^2_0}) of the Weyl tensor squared part we finally have
\begin{widetext}
    \begin{align}
    &\hspace{-0.2cm}\sqrt{g}\,\big\langle T^\alpha_\beta \big\rangle\Big|_{\,\bar g}^{\,g} = -\frac\alpha{4\pi^2}\sqrt{\bar g}\,\big(\bar R^{\mu\nu} +2\bar\nabla^{(\mu}\bar\nabla^{\nu)}\big)\big(\sigma \bar C^\alpha{}_{\mu\beta\nu}\big) 
    +\frac{1}{8\pi^2}\sqrt{g}\,\Big[\,\beta \,^{(3)}\!H^\alpha_\beta
    +\frac\alpha{18}\,^{(1)}\!H^\alpha_\beta+2\beta R^{\mu\nu}C^\alpha{}_{\mu\beta\nu}\,\Big]_{\,\bar g}^{\,g}. \label{BC_gen}
    \end{align}
\end{widetext}
This is a generalization of the Brown--Cassidy formula to the case of a nonzero Weyl tensor. The first term of this expression is Weyl invariant in view of the above remark and can be represented by its unbarred version.

The check of consistency of this formula with the original expression for the conformal anomaly is trivial in view of $^{(3)}\!H^\alpha_\alpha=(E-C^2)/2$, $^{(1)}\!H^\alpha_\alpha=6\Box R$ and tracelessness of the Weyl tensor,
\begin{equation}
\hspace{-0.2cm}\sqrt{g}\,\big\langle T^\alpha_\alpha \big\rangle\Big|_{\,\bar g}^{\,g} = \frac{\sqrt{g}}{16\pi^2}\!\Big[\beta E-\beta C^2+\tfrac{2\alpha}3 \Box R\Big]_{\bar g}^g = -\frac{\sqrt{g}\,a_2}{16\pi^2}\Big|_{\,\bar g}^{\,g},
\end{equation}
where the last equality follows from the conformal invariance of the density $\sqrt{g}\,C^2$ and from the relation \eqref{gamma_vs_alpha} between the coefficients $\gamma$ and $\alpha$, $\alpha=\tfrac32\gamma$.

The recovery of \eqref{BC_gen} from the direct variation of the Wess--Zumino action \eqref{var_WZ} goes as follows. We use metric variational formulae
    \begin{align}
    &\frac{\delta}{\delta g_{\alpha\beta}}\!\int d^4x\,\sqrt{g}\,C^2\sigma = -2\sqrt{g} \big(R_{\mu\nu}\!+2\nabla_{(\mu}\!\nabla_{\nu)}\big) \big(\sigma C^{\alpha\mu\beta\nu}\big),               \label{var_C^2_sigma} \\
    &\frac{\delta}{\delta g_{\alpha\beta}} \int d^4x\,\sqrt{g}\,\calE_4\sigma = \sqrt{g}\,\Delta^{\alpha\beta}\sigma, \\
    &\frac{\delta}{\delta g_{\alpha\beta}} \int d^4x\,\sqrt{g}\,\varphi\Delta_4\sigma=-\frac{\sqrt{g}}2 D^{\alpha\beta}[\varphi,\sigma],
    \end{align}
which hold for generic scalar test functions $\sigma$ and $\varphi$ with the differential operator $\Delta^{\alpha\beta}$ acting on $\sigma$,
\begin{align} \label{TPhi}
\Delta_{\alpha\beta} &= \frac13(g_{\alpha\beta}\Box-\nabla_\alpha\nabla_\beta)\,\Box \nonumber \\
&+\left[\,2(g_{\alpha\beta}P_{\mu\nu} - g_{\alpha\mu}P_{\beta\nu} - g_{\alpha\nu}P_{\beta\mu})+ \frac83g_{\mu\nu}P_{\alpha\beta}\right. \nonumber\\
& \left.+ 2Pg_{\alpha\mu}g_{\beta\nu} - \frac53Pg_{\alpha\beta}g_{\mu\nu} - 2W_{\alpha\mu\beta\nu}\right]\nabla^\mu\nabla^\nu\nonumber\\
&+ \big(\,g_{\alpha\beta}g_{\mu\nu} - g_{\alpha\mu}g_{\beta\nu} - g_{\alpha\nu}g_{\beta\mu}\big) (\nabla^\mu P)\nabla^\nu,
\end{align}
and the bilinear form $D^{\alpha\beta}(\varphi,\sigma)$,
\begin{align} \label{TPhiSigma}
D_{\alpha\beta}[\varphi, \sigma] &= -\frac12g_{\alpha\beta}\Box\varphi\,\Box\sigma - 2\sigma_{\alpha\beta}\Box\varphi \nonumber\\
&+ 2\sigma_\alpha\Box\varphi_\beta - \frac13 g_{\alpha\beta}\sigma_\mu\Box\varphi^\mu - \frac23\varphi_{\mu(\alpha\beta)}\sigma^\mu
\nonumber\\
&+ \left[\, 2W_{\alpha\mu\beta\nu} + \frac13(g_{\mu\nu}R_{\alpha\beta} - g_{\alpha\mu}g_{\beta\nu}R)\right]\varphi^{(\mu}\sigma^{\nu)} \nonumber\\
&+ \frac13\big(\,4\varphi_{\alpha\mu}\sigma^\mu_\beta - g_{\alpha\beta}\varphi_{\mu\nu}\sigma^{\mu\nu}\big) + \big(\,\varphi\Leftrightarrow\sigma\,\big),
\end{align}
where $\varphi_\alpha\equiv\nabla_\alpha\varphi$, $\sigma_{\alpha\beta}\equiv\nabla_\beta\nabla_\alpha\sigma$, $\varphi_{\alpha\beta\gamma}\equiv\nabla_\gamma\nabla_\beta\nabla_\alpha\varphi$, etc. Note that the trace of $\Delta^{\alpha\beta}$ coincides with the Paneitz operator, $g_{\alpha\beta}\Delta^{\alpha\beta}=\Delta_4$, which matches with the conformal variation \eqref{basic}, and the bilinear form $D^{\alpha\beta}(\varphi, \sigma)$ is traceless in view of conformal invariance of $\sqrt{g}\Delta_4$.

Using these relations we get from \eqref{var_WZ} and \eqref{RTF}
\begin{align}
&\sqrt{g}\,\big\langle T^\alpha_\beta \big\rangle\,\Big|_{\,\bar g}^{\,g} = -\frac{\alpha}{4\pi^2}\sqrt{g}\big(R^{\mu\nu} +2\nabla^{(\mu}\nabla^{\nu)}\big)\big(\sigma C^\alpha{}_{\mu\beta\nu}\big) \nonumber \\
&+\frac{\sqrt{g}}{8\pi^2}\!\Big(2\beta\Delta^\alpha_\beta\sigma
+\beta D^\alpha_\beta[\sigma,\sigma]\Big) + \sqrt{g}\left(\tfrac{\gamma}{12}
+\tfrac{\beta}{18}\right) {}^{(1)}\!H^\alpha_\beta\,
\Big|_{\,\bar g}^{\,g}.               \label{var_BC_gen}
\end{align}
The term in the first line here coincides with its barred version in \eqref{BC_gen}---this easily follows from the relation \eqref{var_C^2_sigma} where the integrand can be identically replaced by the barred one. The $\frac\gamma{12}{}^{(1)}\!H^\alpha_\beta$ term here matches with the $\frac\alpha{18}{}^{(1)}\!H^\alpha_\beta$ term of \eqref{BC_gen} in view of the relation $\alpha=\tfrac32\gamma$. And finally, the identity holds
    \begin{align}
    \sqrt{g}\,\Big[{}^{(3)}\!H^\alpha_\beta
    +\tfrac1{18}{}^{(1)}\!H^\alpha_\beta
    &+2 R^{\mu\nu}C^\alpha{}_{\mu\beta\nu}\Big]_{\,\bar g}^{\,g}\nonumber\\
    &=\sqrt{g}\Big(
    2\Delta^\alpha_\beta\sigma + D^\alpha_\beta[\sigma,\sigma]\Big),
    \end{align}
which completely reconciles the two expressions \eqref{BC_gen} and \eqref{var_BC_gen} for the stress tensor behavior on the orbit of the conformal group.

\section{Conformally flat spacetime} \label{ConfFlatSect}

The generalization \eqref{BC_gen} of Brown-Cassidy formula to the case of a nonvanishing Weyl tensor might be not very useful, because in the general case not much can be said about $\langle\,T^\alpha_\beta\rangle\,|_{\bar g}$. Therefore we will restrict ourselves with the case of the conformally flat spacetime for which the conformal transformation of the metric can lead to the metric $\bar g_{\mu\nu}$ of flat spacetime, where  $\langle\,\bar T^\alpha_\beta\rangle$ is either zero or can be obtained from flat space physics. Interestingly, in this case the parameter of the conformal transformation $\sigma$ making this transition satisfies the equation
    \begin{equation} \label{flat_sigma}
    \Delta_4\, \sigma = \frac14 \calE_4
    \end{equation}
and in asymptotically flat case with Dirichle boundary conditions has a unique solution \eqref{varSigmaRFT}, $\sigma=\varSigma_{\rm RFT}$. This, apparently not very well known fact, can be proven by using the equation for the conformal transformation of the four-dimensional Schouten tensor \eqref{Schouten} ($g_{\mu\nu}=e^{2\sigma} \bar g_{\mu\nu}$)
\begin{equation}
P_{\mu\nu}-\bar P_{\mu\nu} = -\sigma_{\mu\nu} -\sigma_\mu\sigma_\nu + \frac{1}{2}\sigma_\alpha\sigma^\alpha g_{\mu\nu},
\end{equation}
where $\sigma_\mu\equiv\nabla_\mu\sigma$ and $\sigma_{\mu\nu}\equiv\nabla_\nu\nabla_\mu\sigma$. Assuming that $\bar g_{\mu\nu}$ is flat space metric with $\bar P_{\mu\nu}=0$, differentiating twice and again using this relation to express $P_{\mu\nu}$ in terms of the derivatives of $\sigma$ one has
\begin{multline}
\nabla^\mu\nabla^\nu\left(P_{\mu\nu}+\sigma_{\mu\nu}
+\sigma_\mu\sigma_\nu- \frac{1}{2}\sigma_\alpha\sigma^\alpha g_{\mu\nu}\right) \\
=\Delta_4\sigma-\frac14\calE_4 = 0,
\end{multline}
whence it follows that the conformal invariant metric \eqref{conf_inv} in the RFT gauge \eqref{RFT_gauge} is actually the flat space one when the Weyl tensor is zero
\begin{equation}
\bar R^\alpha_{\;\;\beta\mu\nu}=0,\quad
\bar g_{\mu\nu}=e^{-2\varSigma_{\rm RFT}[\,g\,]}g_{\mu\nu}\big|_{\,C_{\alpha\beta\mu\nu}=0}.
\end{equation}
Note that $\bar g_{\mu\nu}$ here is not automatically diagonal unit matrix $\delta_{\mu\nu}$, because this is the invariant statement which is valid in any coordinate system.

\subsection{Anomaly driven cosmology \label{slih}}

Applications of the conformal anomaly in the cosmological context have a long history, see for example \cite{FHH, Starobinsky_R^2, FT, HHR, Shapiro_etal_cosmology, Prokopec_etal, AlStar-Shapiro_etal}. In particular, cosmology with the Friedman--Robertson--Walker (FRW) metric represents the situation when the anomalous action $\Delta\varGamma[\bar g,\sigma]$ entirely determines the physics of the field model and via effective equations of motion produces a nontrivial back reaction of quantum matter on the dynamical metric background. The most interesting example is, perhaps, the case when $\varGamma[\,\bar g\,]$ in \eqref{RTF} nontrivially contributes to this back reaction effect rather than just serves as an inert flat space background.

This is the spatially closed cosmology driven by a conformal field theory (CFT) from the initial state in the form of a special microcanonical density matrix, which was orginally suggested in \cite{slih} and recently reviewed in \cite{SLIH_review}. With the density matrix defined as the projector on the space of solutions of the Wheeler--DeWitt equations \cite{why,whyBFV} the statistical sum in this model has a representation of the Euclidean quantum gravity (EQG) path integral
    \begin{equation} \label{Z}
    Z = \int D[\,g_{\mu\nu},\phi\,]\; e^{-S[\,g_{\mu\nu},\phi\,]},
    \end{equation}
where integration runs over the metric $g_{\mu\nu}$ and matter fields $\phi$ which are periodic on the Euclidean spacetime of topology $S^1\times S^3$ with the time $\tau$ compactified to a circle $S^1$.

When the classical action $S[\,g_{\mu\nu},\phi\,]$ is dominated by numerous CFT fields $\varPhi$ with their action $S_{C\!F\!T}[\,g_{\mu\nu},\varPhi\,]$, the statistical sum can be approximated by the contribution of the saddle point of this integral. This is the extremum of the total action including the tree-level gravitational Einstein--Hilbert action $S_{EH}[\,g_{\mu\nu}]$ and the effective action $\varGamma[\,g_{\mu\nu}]$ of these CFT fields\footnote{Disregarding the graviton loops can be justified by the domination of conformal fields outnumbering the metric, and retaining the Einstein--Hilbert term obviously follows from the fact that this term with {\em renormalized} gravitational and cosmological constants is anyway induced from the quantum conformal sector.},
    \begin{align}
    \varGamma_{\rm tot}[\,g_{\mu\nu}] &= S_{EH}[\,g_{\mu\nu}] +\varGamma[\,g_{\mu\nu}], \\
    e^{-\varGamma[\,g_{\mu\nu}]} &= \int D\varPhi\, e^{-S_{C\!F\!T}[\,g_{\mu\nu},\varPhi\,]}. \label{GammaCFT}
    \end{align}

Choosing as $g_{\mu\nu}$ the FRW metric with the scale factor $a(\tau)$ and the lapse function $N$ ($\varOmega^2_{(3)}$ is the metric of the  3-dimensional sphere of a unit radius),
\begin{equation}
ds^2=N^2d\tau^2+a^2d\varOmega^2_{(3)}=
a^2(\tau)\big(d\eta^2+d\varOmega^2_{(3)}\big),
\end{equation}
one immediately finds that in terms of the conformal time variable $\eta$, related to the Euclidean time $\tau$ by the relation $d\eta=d\tau/a(\tau)$, this metric is conformally equivalent to the metric $\bar g_{\mu\nu}\equiv g_{\mu\nu}^{ EU}$ of the Einstein static universe with spatial sections---the 3-dimensional spheres of some constant radius $a_0$,
\begin{gather}
d\bar s^2=a_0^2 \big(d\eta^2+d\varOmega^2_{(3)}\big)\equiv
g_{\mu\nu}^{ EU}dx^\mu dx^\nu, \\
ds^2=e^{2\sigma} d\bar s^2, \quad g_{\mu\nu}=e^{2\sigma} g_{\mu\nu}^{ EU},\quad \sigma=\ln\frac{a}{a_0}.
\end{gather}
Therefore the CFT effective action expresses in terms of the same action on a static Einstein universe $\varGamma[\,g_{\mu\nu}^{ EU}]\equiv\varGamma_{ EU}$ and Wess--Zumino action \eqref{RTF} with the above conformal parameter $\sigma$
\begin{equation}
\varGamma[\,g_{\mu\nu}]=
\Delta\varGamma[\,g_{\mu\nu}^{ EU},\sigma\,]
+\varGamma_{EU}.                   \label{delta_Gamma_EU}
\end{equation}

The calculation of $\varGamma_{EU}$ is strongly facilitated by the static nature of the background, but it still yields a nontrivial result in view of compactification of time on $S^1$. To begin with, note that although $g_{\mu\nu}^{ EU}$ explicitly depends on the size $a_0$ of $S^3$, the value of $\varGamma_{EU}$ is $a_0$-independent for a fixed period of the conformal time $\eta=\oint d\eta$. This follows from the invariance of the effective action under global conformal transformations \eqref{global} for conformally flat spacetimes with zero bulk part of the Euler characteristics (which is the case of $S^1\times S^3$). This also can be confirmed by using scaling properties of the conformal fields. Indeed, the energies of conformal quanta on a static spacetime scale as $1/a_0$ and their Hamiltonian reads,
\begin{equation}
\hat H=\sum_\omega\frac\omega{a_0}\Big(\hat a^\dagger_\omega\hat a_\omega\pm\frac12\Big),
\end{equation}
where summation runs over all quantum numbers (and spins) of the energies $\omega/a_0$ of all field oscillator modes on a static 3-dimensional sphere of the radius $a_0$ and $\hat a^\dagger_\omega$ and $\hat a_\omega$ are the relevant creation-annihilation operators ($\pm$ signs correspond to bosons or fermions). The path integral over (anti)periodic conformal (fermion) boson fields with a period ${\cal T}=\oint d\tau N$ on a static metric background is exactly calculable and equals the equilibrium statistical sum at the temperature $1/{\cal T}$ which expresses as a function of the conformal time period $\eta={\cal T}/a_0$
\begin{multline}
e^{-\varGamma_{ EU}}=\int D\varPhi\,
    e^{-S_{C\!F\!T}[\,g_{\mu\nu}^{ EU},\varPhi\,]} \\
    ={\rm Tr}\,e^{-{\cal T}\hat H}
    =\exp\big(-\eta E_{\rm vac}-F(\eta)\big).  \label{varGamma_EU}
\end{multline}
Here $F(\eta)$ is the free energy of the gas of conformal particles and $E_{\rm vac}$ is a UV divergent Casimir energy which should be covariantly renormalized
    \begin{align}
    F(\eta) &= \sum_{\omega}\Big[\pm\,\ln\big(1\mp e^{-\omega\eta}\big)\,\Big],\\
    E_{\rm vac} &= \Big(\sum_\omega\frac{\pm\,\omega}2\Big)_{\rm ren}.                  \label{E_vac}
    \end{align}
Thus, the dependence on $a_0$ is absorbed into the dependence on $\eta$ which should be fixed under the rescaling of $a_0$. Note that it is $\eta$ that should be kept fixed under the global conformal transformation which simultaneously rescales the lapse function $N$ and $a_0$ in the definition of the conformally invariant $\eta=\oint d\tau\,N/a_0$.

Remarkably, the covariant renormalization of the vacuum Casimir energy $E_{\rm vac}$ also follows from the behavior of the effective action on the orbit of the conformal group. The Einstein universe extending from $-\infty$ to $+\infty$  in $\eta$ is mapped to flat space by the transition to the radial coordinate $\rho$
\begin{equation}
\eta\mapsto \rho = a_0 e^\eta, \quad
-\infty<\eta<+\infty,\quad
0\leq\rho<\infty,                 \label{ES1}
\end{equation}
with the conformal relation between the two metrics
\begin{align}
&ds^2_{EU}=e^{2\sigma} ds_{\rm flat}^2, \quad \sigma = -\eta=\ln\frac{a_0}{\rho},  \label{sigma_FRW}\\
&ds_{\rm flat}^2=d\rho^2+\rho^2 d\varOmega^2_{(3)}.
\end{align}
For the vacuum state (the limit $\eta\to\infty$ and $F(\eta)\to 0$ in Eq.~\eqref{varGamma_EU}) $\varGamma_{ EU}\to E_{\rm vac}\eta$. On the other hand, from Eq.~\eqref{RTF} with the above expression for $\sigma$
\begin{align}
&\Delta\varGamma[\,g_{\rm flat},\sigma\,]=\frac\beta{8\pi^2}\int d^4x\sqrt{g_{\rm flat}}\big(\Box_{\rm flat}\sigma\big)^2\nonumber\\
&\qquad\qquad\quad-\frac{1}{32\pi^2} \Big(\frac\gamma6 + \frac{\beta}{9}\Big)\,\int d^4x\sqrt{g_{EU}}R^2_{EU}.
\end{align}
Bearing in mind that $\Box_{\rm flat}\sigma=-2/\rho^2$, $\int d^4x\sqrt{g_{\rm flat}}\!\mapsto 2\pi^2\int d\rho\,\rho^3$, $R_{EU}=6/a_0^2$ and $\int d^4x\sqrt{g_{EU}}\mapsto 2\pi^2a_0^4\int d\eta$, one has
\begin{align}
\hspace{-0.2cm}\varGamma_{ EU}-\varGamma_{\rm flat}=\Delta\varGamma[\,g_{\rm flat},\sigma\,]&=\beta\!\int\frac{d\rho}\rho-
\Big(\frac38 \gamma + \frac{\beta}{4}\Big)\!\int d\eta\nonumber\\ &=\frac34\,\Big(\beta-\frac\gamma2\Big)\int d\eta.
\end{align}
Therefore, under an obvious assumption that $\varGamma_{\rm flat}=0$ one has
\begin{equation}
E_{\rm vac}=\frac34\,\Big(\beta-\frac\gamma2\Big). \label{Casimir}
\end{equation}
In other words, after covariant renormalization by covariant counterterms the Casimir energy gets the value compatible with the behavior of the renormalized effective action on the conformal group orbit (or with the Brown--Cassidy formula for the vacuum stress tensor). This compatibility was indeed checked by direct renormalization of the UV divergent sum over field modes in \eqref{E_vac} \cite{Dowker-Critchley, Mamaev-Mostepanenko-Starobinsky, Ford, Candelas-Dowker}.

Let us now turn to the contribution of the conformal transformation from the generic FRW metric to that of the static Einstein universe in \eqref{delta_Gamma_EU}. To begin with we use the freedom of finite renormalization \eqref{renormaction1} which reduces the theory to the case of anomaly \eqref{anomaly} with $\gamma=0$ and, in particular, renders $E_{\rm vac}=\tfrac34\beta$. In the cosmological context this freedom corresponds to the adjustment of the coupling constant of the Starobinsky $R^2$-action \cite{Starobinsky_R^2} which plays an important role in inflation theory and the dark energy model. Then, with $\gamma=0$ and $\sigma$ given by \eqref{sigma_FRW} the Wess--Zumino term in \eqref{delta_Gamma_EU} takes the form \cite{slih}
    \begin{equation}
    \varGamma_{\rm Ren}[\,g\,]-\varGamma_{\rm Ren}[\,g_{ EU}] = \frac{3\beta}2\oint d\tau\,N \left(\frac{a'^2}{a} - \frac{a'^4}{6\,a}\right),         \label{delta_Gamma_R}
    \end{equation}
when written down in terms of the original FRW coordinates with the notation for the invariant time derivative $a'=da/Nd\tau$. Note that the result is again independent of the constant $a_0$ because it contains only differentiated $\sigma$ and, moreover, it does not involve higher order derivatives of $a(\tau)$. The last property is entirely due to the fact of $\gamma$ being renormalized to zero and due to the cancellation of higher derivative terms in the minimal form of Wess-Zumino action \eqref{minimal}.

Now we assemble together the Einstein-Hilbert action (with the reduced Planck mass $M_{\rm P}=1/\sqrt{8\pi G}$ and the cosmological constant $\varLambda$), the action on the Einstein universe space \eqref{varGamma_EU} and \eqref{delta_Gamma_R}. This leads to the total effective action on the generic Euclidean FRW background periodic in Euclidean time with the period $\eta$ measured in units of the conformal time
    \begin{align}
    &\varGamma_{\rm tot}[\,a,N\,] = 6\pi^2 M_P^2\oint d\tau\,N \bigg\{-aa'^2
    -a+\frac\varLambda3 a^3\nonumber\\
    &+\frac{\beta}{4\pi^2 M_P^2}\left(\frac{a'^2}{a} -\frac{a'^4}{6 a} +\frac1{2a}\right)\bigg\} + F(\eta),                           \label{effaction0}\\
    &\eta=\oint\frac{d\tau N}a.     \label{period}
    \end{align}
Here the contribution of the conformal anomaly and Casimir energy \eqref{Casimir} (with $\gamma=0$) are both weighted by the parameter $\beta$ of the topological term in the conformal anomaly. The free energy of the gas of conformal particles $F(\eta)$ is a function of the effective (``comoving'') temperature of this gas -- the inverse of the circumference $\eta$ of the cosmological instanton (\ref{period}). Despite essentially non-stationary metric background this gas stays in equilibrium state because of scaling properties of its particles and produces back reaction on the Friedmann metric background.

Applications of the action (\ref{effaction0}) have been considered in the number of papers \cite{slih,slih_R^2,hill-top} and recently reviewed in \cite{SLIH_review}. Physics of the CFT driven cosmology is entirely determined by this effective action and the effective (Euclidean) Friedmann equation. The latter follows from the action by varying the lapse $N(\tau)$ and expressing the Hubble factor in terms of the energy density. In cosmic type gauge $N=1$, $\dot a=da/d\tau$, it reads
    \begin{align}
    &\frac1{a^2}-\frac{\dot a^2}{a^2}=\frac{\varepsilon}{3M_\pm^2\big(\varepsilon\big)},\\
    &\varepsilon=M_P^2\varLambda
    +\frac1{2\pi^2 a^4}\sum_\omega\frac\omega{e^{\eta\omega}-1},\\
    &M_\pm^2(\varepsilon)=\frac{M_P^2}2\Big(1\pm\sqrt{1
    -\beta\varepsilon/6\pi^2M_P^4}\,\Big),
    \end{align}
where the total energy density $\varepsilon$ includes the cosmological constant contribution and the radiation density of conformal field modes distributed over Planckian spectrum with the comoving temperature $1/\eta$. The nonlinear effect of the Weyl anomaly manifests itself in the effective Planck mass squared explicitly depending on $\varepsilon$ which takes two possible values $M_\pm^2(\varepsilon)$.\footnote{To avoid mixup of the signs in $M_\pm^2$ and sign factors associated with the statistics of conformal $\omega$-modes we present here the radiation spectrum only for bosonic case.}  These equations should be amended by the expression for the conformal time period that interpolates between the turning points of the solution with $\dot a(\tau)=0$. Note that the right hand side of the Friedmann equation does not contain Casimir energy density -- it turns out to be fully screened due to the dynamical effect of the Weyl anomaly. This is the result of the finite renormalization (\ref{renormaction1}) leading to a particular value of the anomaly coefficient of $\Box R$, $\gamma=0$.

For the choice of $+$ sign in $M_\pm^2$ the solutions of this quantum Friedmann equation turn out to be the so-called {\em garlands} -- the cosmological instantons of $S^1\times S^3$ topology, which have the periodic scale factor $a(\tau)$  oscillating on $S^1$ between maximal and minimal values $a_\pm$ \cite{slih}. These instantons serve as initial conditions for the cosmological evolution in the physical Lorentzian spacetime. This evolution follows from $a(\tau)$ by the analytic continuation $a_L(t)=a(\tau_++it)$, $(da_L/dt)^2=-\dot a^2$, to the complex plane of the Euclidean time at the turning point with the maximal scale factor $a_+=a(\tau_+)$. It can incorporate a finite inflationary stage if the model is generalized to the case when a primordial cosmological constant is replaced by the potential of the inflaton field $\phi$, $\varLambda\to V(\phi)/M_P^2$, staying in the slow-roll regime during the inflationary stage\footnote{Alternatively, the role of inflaton can be played by Ricci curvature in the Starobinsky $R^2$-model, the coupling of the $R^2$ term being subject to the renormalization respecting the zero value of $\alpha$ in the total Weyl anomaly \cite{slih_R^2}.} and decaying in the end of inflation by a usual exit scenario \cite{slih_R^2,hill-top}. The energy scale of inflation -- its Hubble parameter $H\sim\sqrt{\varLambda/3}$ turns out to be bounded from above by $\sqrt2\pi M_P/\sqrt\beta$, so that to solve the problem of hierarchy between the Planck and inflation scales one needs $\beta\gg 1$ which matches with the previously adopted assumption that numerous conformal fields drastically outnumber all other fields and dominate over their loop corrections.

For the negative sign in $M_\pm^2$ the solutions represent vacuum $S^4$-instantons of the no-boundary type with the vanishing minimal value of the scale factor $a_-=0$. They correspond to the diverging $\eta\sim\int_0^{a_+}da/a\dot a\to \infty$ or zero temperature. These solutions, however, do not contribute to the statistical sum because of their infinitely positive action $\varGamma_{\rm tot}\to+\infty$ ---  the quantum effect of the trace anomaly which flips the sign of the negative tree-level action of the Hartle-Hawking instantons \cite{HH} and sends it to $+\infty$ \cite{slih}. Thus the CFT cosmology scenario is free from the infrared catastrophe of the no-boundary quantum state which would imply that the origin of an infinitely big Universe is infinitely more probable than that of a finite one.

\section{Renormalization group and the metamorphosis of the running scale}
This section has essentially discussion nature and is associated with the covariant perturbation theory of the above type. One of the motivations for this discussion is that, in spite of a widespread concept of running cosmological and gravitational constants, which is especially popular within the asymptotic safety approach, there is a very profound and persuading criticism of this concept \cite{Donoghue}. It is based on numerous arguments of the tadpole structure of the cosmological and Einstein terms, on concrete results for graviton scattering amplitudes \cite{Donoghue_grav_const} which cannot be interpreted in terms of a universal scaling of $\varLambda$ and $G$, etc.

At the same time in renormalizable gravity models with multiple couplings the solution of the full set of RG equations includes running cosmological and gravitational constants \cite{Tseytlin_quadratic}. So the question arises how to interpret their running scale. Here is the attempt to do this in terms of the covariant curvature expansion developed in \cite{CPTI,CPTII, CPTIII}.

We start with the classical action which is the sum of local curvature invariants of growing dimensionality $(4+m)$ in units of the mass
\begin{equation}
S[\,g_{\mu\nu}]=\sum\limits_{m,N}\varLambda^{(m)}_N\int d^4x\,\sqrt{g}\,\Re^{(4+m)}_N(x).             \label{class_action}
\end{equation}
They are monomials of $N$-th order in curvature tensors which are acted upon by covariant derivatives
\begin{eqnarray}
&&\Re^{(m)}_N(x)=\mathop{\underbrace{\nabla...\nabla}_{m-2N}}
\mathop{\overbrace{\Re(x)...\Re(x)}^{N}},\label{local_Re}\\
&&{\rm dim}\; \Re^{(m)}_N(x)
\equiv \big[\,\Re^{(m)}_N(x)\,\big]=m.
\end{eqnarray}
The curvature monomials enter the action with coupling constants $\varLambda^{(m)}_N$ of the decreasing (with growing $m$) dimensionality
\begin{equation}
[\,\varLambda^{(m)}_N\,]=d-m, \quad m=0,1, \ldots .  \label{couplings}
\end{equation}
Summation in \eqref{class_action} can run over finite set of terms providing the renormalizability of the theory, or formally extended to the infinite set in the framework of generalized RG theory with infinite set of couplings $\{\varLambda\}=\varLambda^{(m)}_N$.

Within covariant perturbation theory the full metric is decomposed as a sum of the flat spacetime metric $\tilde g_{\mu\nu}$ and the perturbation $h_{\mu\nu}$
\begin{equation}
g_{\mu\nu}=\tilde g_{\mu\nu}+h_{\mu\nu},
\end{equation}
so that each curvature invariant becomes expanded as an infinite series in powers of $h_{\mu\nu}$ forming a new set of $h$-monomials on the flat space background
\begin{eqnarray}
&&\int d^4x\,\sqrt{g}\,\Re^{(m)}_N=\sum\limits_{M=N}^\infty
\int d^4x\, \sqrt{\tilde g}\,I_M^{(m)}(h),\nonumber\\
&&I_M^{(m)}(h)\propto \mathop{\underbrace{\tilde\nabla...\tilde\nabla}_{m}} \mathop{\overbrace{h(x)...h(x)}^{M}}.              \label{invariant}
\end{eqnarray}

Then in the notations of the covariant perturbation theory the calculation of the  renormalized effective action leads to the same sequence of monomials acted upon by the operator form factors $\varGamma_n^{(i)}\big(\{\varLambda\}, \tilde\nabla_1,...\tilde\nabla_1\big)$ which make them nonlocal, $\{\varLambda\}$ denoting the full set of couplings (\ref{couplings}). Within dimensional regularization these renormalized coupling constants get rescaled by the normalization parameter $\mu$ and expressed in terms of their dimensionless analogues $\lambda^{(m)}_N(\mu)$
\begin{equation}
\varLambda^{(m)}_N=\mu^{d-m}\lambda^{(m)}_N(\mu),
\end{equation}
and the perturbation theory form factors also express as the functions of dimensionless arguments
\begin{equation}
\varGamma_M^{(m)}\big(\{\varLambda\},\tilde\nabla_1,...\tilde\nabla_M\big)=
\mu^{d-m}\gamma_M^{(m)}\Big(\{\lambda(\mu)\},\tfrac{\tilde\nabla_1}\mu,...
\tfrac{\tilde\nabla_M}\mu\Big)
\end{equation}
 Correspondingly the effective action becomes
\begin{align}
&\hspace{-0.2cm}\varGamma[\,g_{\mu\nu}]=\sum\limits_{(m)}\mu^{d-m}\sum\limits_{M=0}^\infty
\int d^dx\, \sqrt{\tilde g} \nonumber\\ &\times\gamma_M^{(m)}\big(\{\lambda(\mu)\},\tfrac{\tilde\nabla_1}\mu,...
\tfrac{\tilde\nabla_M}\mu\big)
I_M^{(m)}(h_1,h_2,...h_M)\,\big|_{\,\{x\}=x},  \label{h_action}
\end{align}
where $I_M^{(m)}(h_1,h_2,...h_M)$ is the analogue of the invariant (\ref{invariant}) with split spacetime arguments. A typical assumption of the RG theory that the renormalized action is independent of the running scale then leads to the set of equations for $\lambda^{(m)}_N(\mu)$ with the beta functions following from the residues of spacetime dimension poles in the formfactors $\varGamma_M^{(m)}\big(\{\lambda(\mu)\},\{\tilde\nabla/\mu\}\big)$,
\begin{equation}
\mu\frac{d}{d\mu}\varGamma[\,g_{\mu\nu}]=0 \to \mu\frac{d}{d\mu}\lambda^{(m)}_N(\mu)=\beta^{(m)}_N(\mu)
\big(\{\lambda(\mu)\}\big).
\end{equation}

A critical step now consists in the choice of the running scale which could probe the high energy limit of the theory and embrace a simultaneous scaling of all formfactors and invariant monomials of (\ref{h_action}). Then the replacement of the parameter $\mu$ by this scale will identically bring the effective action to the form explicitly revealing its UV limit. The choice of this scaling object can be very different depending on the concrete physical setup. If the theory has a dimensional scalar field $\phi$ with a nonvanishing and slowly varying mean value it would be natural to identify RG normalization $\mu$ with $\phi$. This would lead to the nontrivially ``running'' in $\phi$ of the cosmological and Einstein terms, $\varLambda\to\varLambda(\phi)$ and $G\to G(\phi)$, (amended of course by a gradient expansion series in derivatives of $\phi$), but of course these terms acquire the interpretation of the Coleman-Weinberg type potential and nonminimal coupling of $\phi$ to the scalar curvature.

We, however, are interested in the UV scaling of all derivatives $\tilde\nabla\to\infty$, which in momentum space representation of scattering amplitudes is conventionally represented by the high energy Mandelstam invariants or some other combinations of external momenta. In the coordinate representation of the covariant perturbation theory of \cite{CPTI,CPTII,CPTIII} the role of this scale should be played by some operator. So we suggest as a candidate for this object the following nonlocal operator $\tilde D$ which also formally tends to infinity in the limit of $\tilde\nabla\to\infty$ and in fact embraces a simultaneous scaling of all invariant monomials in (\ref{h_action}),
\begin{equation}
\tilde D\equiv\Big(-\sum_{N=1}^\infty
\tilde\Box_N\Big)^{1/2}, \quad \tilde\Box_N\equiv \tilde g^{\mu\nu}\tilde\nabla_\mu\tilde\nabla_\nu.             \label{tilde_D}
\end{equation}
Though being very formal, this operator is well defined in each $N$-th monomial order because it becomes truncated to the finite sum when acting on the monomial of $N$ perturbations $h_1,...h_N$, and for $N=0$ it is just zero because of its action on an independent of $x$ constant,
\begin{equation}
\tilde D_N\equiv\bigg(-\sum_{M=1}^N
\tilde\Box_M\bigg)^{1/2}, \quad \tilde D_0=0. \label{tilde_D_N}
\end{equation}

In the UV domain $\tilde\nabla_n\to\infty$, when $\tilde\nabla_n/\tilde D_N=O(1)$, $n\leq N$, the formfactors in each $N$-th order become after the replacement $\mu\to \tilde D$ the functions of a single operator variable $\tilde D_N$,
\begin{align}
&\hspace{-0.1cm}\mu^{4-m}\gamma_N^{(m)}\big(\lambda(\mu)\,
\big|\,\tfrac{\tilde\nabla_1}\mu,...
\tfrac{\tilde\nabla_N}\mu\big)\,\Big|_{\,\mu\to\tilde D_N}\;\to\nonumber\\
&\;(\tilde D_N)^{4-m}\gamma_N^{(m)}\big(\lambda(\tilde D_N)\,\big|\,O(1)\big)\equiv
(\tilde D_N)^{4-m}\lambda_N^{(m)}\big(\tilde D_N\big), \label{scaling}
\end{align}
and the expansion of the formally independent of $\mu$ action takes the form
\begin{align}
&\varGamma[\,g_{\mu\nu}]\,\Big|_{\,\mu\to\tilde D}\to\,\sum\limits_{m}\sum\limits_{N=0}^\infty
\int d^4x\, \sqrt{\tilde g}\nonumber\\
&\times\,(\tilde D_N)^{4-m}
\lambda_N^{(m)}(\tilde D_N)\,
I_N^{(m)}(h_1,h_2,...h_N)\,\Big|_{\,\{x\}=x}.    \label{h_expansion}
\end{align}

The next step consists in the recovery of the covariant form of the expansion in terms of the original spacetime curvature. Curiously, despite the fact that the covariant perturbation theory of \cite{CPTI,CPTII,CPTIII} is rather often being referred to in literature, subtle details of this step are usually disregarded which leads to confusing statements on the ambiguity of this procedure, dependence on the gauge by which the metric perturbation $h_{\mu\nu}$ is related to the curvature \cite{Gorbar-Shapiro_running}, etc. At the same time, this procedure is unique, provided that one does not treat $\tilde g_{\mu\nu}$ and $\tilde\nabla_\mu$ as Cartesian $\delta_{\mu\nu}$ and $\partial_\mu$, but rather proceeds in generic coordinate system and uses the only invariant statements that the curvature of the tilded metric is vanishing $\tilde R^\alpha_{\;\;\beta\mu\nu}=0$. This is the covariant equation for $\tilde g_{\mu\nu}$ in terms of the curved metric $g_{\mu\nu}$ and its curvature $R^\alpha_{\;\;\beta\mu\nu}$, whose solution exists as perturbation expansion in $R^\alpha_{\;\;\beta\mu\nu}$ and also requires imposing the gauge \cite{CPTI,CPTII}. But the result of substituting this solution back into manifestly noncovariant (double field) series (\ref{invariant}) is gauge independent because of the implicit invariance of the left hand side of (\ref{invariant}).

In the convenient DeWitt type gauge $\tilde\nabla^\nu h_{\mu\nu}-\tfrac12\nabla_\mu h=O[\,h^2]$, $h\equiv\tilde g^{\alpha\beta}h_{\alpha\beta}$, the solution for $h_{\mu\nu}$ and $\tilde\nabla_\mu$ in terms of $g_{\mu\nu}$ and $\nabla_\mu$ reads in the lowest order as \cite{CPTI,CPTII}
\begin{equation}
h_{\mu\nu}=-\frac2{\Box}R_{\mu\nu}
+O[\,\Re^2],\quad
\tilde\nabla_\mu=\nabla_\mu+O[\,\Re\,].  \label{h_vs_R}
\end{equation}
Using this in (\ref{h_expansion}) we get the replacement of $h$-monomials by the covariant curvature monomials along with the replacement of $\tilde D_N$ by $D_N$,
\begin{align}
&I_N^{(m)}(h_1,h_2,...h_N)\to\nonumber\\ &\qquad\quad\frac1{\Box_1...\Box_N}\Re^{(m+2N)}_N(x_1,...x_N)+O[\,\Re^{N+1}],\\
&\tilde D_N\to D_N+O[\,\Re\,],
\end{align}
where $D_N$ is obviously defined by \eqref{tilde_D_N} in terms of full-fledged covariant d'Alembertians $\Box=g^{\mu\nu}\nabla_\mu\nabla_\nu$, and we reabsorb the coefficient $(-2)^n$ into the symbolic definition of the $N$-th order covariant monomial -- the analogue of the local $\Re^{(m)}_N(x)$, see Eq.~\eqref{local_Re}, with split $N$ spacetime arguments
    \begin{equation}
    \hspace{-0.2cm}\Re^{(m)}_N(x_1,...x_N)=
    \mathop{\underbrace{\nabla...\nabla}_{m-2N}}
    \Re(x_1)...\Re(x_N),\, N\geq 1.                \label{Re_m}
    \end{equation}
For $N=0$ this monomial can be defined as an irrelevant constant bringing no contribution in the UV limit.

Thus the UV limit of the effective action takes the form
\begin{align}
&\hspace{-0.3cm}\varGamma[\,g_{\mu\nu}]\to\int d^4x\,\sqrt{g}
\sum\limits_{m,N\geq 0}^\infty\!
\frac{\lambda_N^{(m)}(D_N)(D_N)^{4-m}}{\Box_1...\Box_N}\nonumber\\
&\qquad\qquad\qquad\times\,
\Re^{(m+2N)}_N(x_1,\cdots x_N)\,\Big|_{\,\{x\}=x},  \label{sum_m}
\end{align}
where we remind that the dimensionless formfactors $\lambda_N^{(m)}(D_N)$ follow from the running RG couplings of the theory $\lambda_N^{(m)}(\mu)$ by the replacement of $\mu$ with the operator $D_N$.

Let us consider application of this result to the cosmological constant sector involving the metric invariants of dimensionality  $m=0$ and $\varLambda^{(4)}_0=\varLambda/16\pi G$. This classical cosmological term gives rise to the infinite set of zero dimension invariants
\begin{align}
&\int d^4x\,\sqrt{g}=\sum\limits_{n=0}^\infty
\int d^4x\,\sqrt{\tilde g}\,I_n^{(0)}(\tilde g,h),    \label{m=0_exp}\\
&\;\;I_0^{(0)}(\tilde g,h)=1,\;\;
I_1^{(0)}(\tilde g,h)=-\tfrac12h,\nonumber\\
&\;\;I_2^{(0)}(\tilde g,h)=\tfrac14h^2-\tfrac12 h_{\mu\nu}^2, \ldots
\end{align}
(indices are contracted by the flat metric and $h=\tilde g^{\mu\nu}h_{\mu\nu}$), whereas at the quantum level they generate the sequence of high energy $m=0$ structures of (\ref{sum_m})
\begin{equation}
\int\! d^4x\sqrt{g}
\sum\limits_{N=2}^\infty \lambda_N^{(0)}(D_N)
\frac{(D_N)^{4}}{\Box_1...\Box_N}\,\Re^{(2N)}_N(x_1,...\, x_N)\Big|_{\{x\}=x},
\end{equation}
where the zeroth order term is zero in view of $D_0=0$ (see Eq.(\ref{tilde_D}) and the first order term is also absent due to its tadpole (total derivative) nature -- remember that $D_1=(-\Box_1)^{1/2}$ and $D_1^4/\Box_1=\Box_1$ is acting on $\Re^{(2)}_1(x_1)$.\footnote{Important caveat is necessary here concerning the annihilation of the total derivative terms. The surface terms at infinity should be vanishing, which is equivalent to a good IR behavior of the nonlocal form factor $\lambda_1^{(0)}(D_1)$ at $\Box\to 0$. We will assume this property basing on the maximum logarithmic singularity of $\lambda_1^{(0)}(D_1)$ which is a function of $\log(-\Box)$ solving the RG equation. The same also applies to integrations by parts considered in what follows. Otherwise, the procedure of subtracting the boundary terms, like the Gibbons-Hawking surface action at asymptotically flat infinity, will be needed, which we briefly discuss below.}

The expansion starts at $N=2$ with the term which has the following structure
\begin{align}
&4\sum\int d^4x\,\sqrt{g}\,\Re^{(2)}(x)\,\lambda_2^{(0)}
\big(\sqrt{-2\Box}\big)\,\Re^{(2)}(x)\nonumber\\
&\;\;
=\int d^4x\sqrt{g}\Big(R_{\mu\nu}F_1(\Box)R^{\mu\nu}+
RF_2(\Box)R\Big)\!+\!O[\,\Re^3].                               \label{quadratic}
\end{align}
Here we took into account that the set of invariants $\Re^{(4)}_2(x_1,x_2)$ can be represented as a sum of terms factored out into the products of Ricci tensors and Ricci scalars with some coefficients\footnote{Bilinear in Riemann curvature terms under the integration sign also reduce to bilinear combinations of $R_{\mu\nu}$ and $R$ by using the expression for Riemann tensor in terms of the Ricci one \cite{CPTI,CPTII}, see footnote \ref{Riemann}.} $a$ and $b$,
\begin{equation}
\Re^{(4)}_2(x_1,x_2)=aR_{\mu\nu}(x_1)\,R^{\mu\nu}(x_2)+b R(x_1)R(x_2),
\end{equation}
and also used an obvious corollary of integration by parts
\begin{align}
&\int d^4x\sqrt{g}\,F(\Box_1,\Box_2)\Re(x_1)\Re(x_2)\,
\Big|_{\,\{x\}=x}\nonumber\\
&\qquad\qquad=\int d^4x\sqrt{g}\,\Re(x)F(\Box,\Box)\Re(x).   \label{int_by_parts}
\end{align}

Remarkable feature of the expression (\ref{quadratic}) is that the power-law operator factors in $(D_N)^{4}/\Box_1...\Box_N$ at $N=2$ completely cancelled out to give the dimensionless formfactors  $F_1(\Box)$ and $F_2(\Box)$ which originate as linear combinations of relevant running $\lambda_2^{(0)}
\big(\sqrt{-2\Box}\big)$ obtained by solving the RG equation. Even more remarkable is the fact that this is a nonlocal term which is quadratic in the curvature even though it has originated from the sector of cosmological term expanded in the series of zero dimension invariants. This is what can be called as {\em metamorphosis to high-energy partners} of the cosmological constant suggested by J.Donoghue in \cite{Donoghue_partners}. Their structure is a direct corollary of the dimensionality arguments within RG approach. The arising form factors of the curvature squared terms are the descendants of RG running couplings of the zero dimension invariants which participate in the decomposition of the cosmological constant term.

In fact, the same structure (\ref{quadratic}) gets reproduced for the contribution of any dimension $m$ in the expansion (\ref{sum_m}). For even dimensionality\footnote{For the set of 2-dimensional curvatures $\Re$ only even dimensions $m$ enter the expansion (\ref{sum_m}), but this can always be generalized to the case of odd-dimensional ``curvatures'', like for example the extrinsic curvature in Ho\v rava gravity models.}, $m\to 2m$, this can be easily demonstrated by decomposing any $(2m+4)$-dimensional quadratic invariant as this was done above
\begin{equation}
\hspace{-0.3cm}\Re^{(2m+4)}_2(x_1,x_2)=\!\!\!\!
\sum_{m_1+m_2=2m}\!\!\!\!\Re^{(m_1+2)}_1(x_1)\Re^{(m_2+2)}_1(x_2).
\end{equation}
Using this in (\ref{sum_m}) one has complete cancellation of the dimensional factor $(D_2)^{4-2m}/\Box^2\sim \Box^{-m}$ in the expression
\begin{align}
&\hspace{-0.3cm}\int d^4x\,\sqrt{g}\!
\sum\limits_{m_1+m_2=2m}\Re^{(m_1+2)}_1(x)\nonumber\\
&\hspace{-0.5cm}\qquad\times\frac{\lambda_2^{(2m)}(D_2)(D_2)^{4-m_1-m_2}}{\Box^2}\,
\Re^{(m_2+2)}_1(x)\nonumber\\
&\hspace{-0.3cm}=\!\int d^4x\sqrt{g}
\Big(R_{\mu\nu}F_1(\Box)R^{\mu\nu}+
RF_2(\Box)R\Big)\!+O[\,\Re^3].          \label{quadratic_G}
\end{align}
Noting that with $\Re^{(m+2)}_1=\mathop{\overbrace{\nabla\cdots\nabla}^{m}} \Re^{(2)}_1$ this follows from integration by parts and the use of various corollaries of contracted Bianchi identity ($\nabla^\nu R_{\mu\nu}=\tfrac12\nabla_\mu R$, etc.),
\begin{align}
&\int d^4x\,\sqrt{g}\!\sum\limits_{m_1+m_2=2m}
\mathop{\underbrace{\nabla\cdots\nabla}_{m_1}}
\Re(x)F(\Box)\mathop{\underbrace{\nabla\cdots\nabla}_{m_2}}\Re(x)\nonumber\\
&=\int d^4x\,\sqrt{g}\Big(R_{\mu\nu}\,\Box^m F_1(\Box)R^{\mu\nu}+
R\,\Box^m F_2(\Box) R\Big)\nonumber\\
&\qquad+O[\,\Re^3]. \label{quadratic_G1}
\end{align}
Here the operators $F_1(\Box)$ and $F_2(\Box)$ have the same dimension as $F(\Box)$ and originate from $F(\Box)$ by the algebra of contracting the indices of covariant derivatives. Using this relation in the left hand side of (\ref{quadratic_G}) one gets the right hand side with completely cancelled powers of $\Box$.

Thus, Eq.(\ref{quadratic_G}) with $m=2$ implies the conversion of the gravitational coupling constant into the dimensionless formfactors of the Einstein term partners. These partners have the same structure as the cosmological term partners quadratic in curvatures. This is again the metamorphosis of RG running of the form $1/16\pi G(\mu)=\mu^2\lambda^{(2)}_2(\mu)\to F_{1,2}(\Box)$.

Note that all this takes place in the UV limit where all curvatures in their monomials are rapidly varying in spacetime with their derivatives $\nabla\to\infty$. At intermediate energies, when the mass scale $M$ surfaces up, the scaling (\ref{scaling}) ceases to make sense and roughly should be replaced with $D\sim M$, and instead of (\ref{quadratic}) one gets exactly the cosmological constant partners of Donoghue \cite{Donoghue_partners} which have the structure of
\begin{equation}
\hspace{-0.2cm}M^4\!\int d^4x\,\sqrt{g}\Big(R_{\mu\nu}
\frac{F_1^{\rm part}(\Box)}{\Box^2}R^{\mu\nu}+
R\,\frac{F_2^{\rm part}(\Box)}{\Box^2}R\Big).   \label{CC_partner}
\end{equation}
The dimensionless form factors $F^{\rm part}_{1,2}(\Box)$ here are accumulating loop corrections with nonlocal logarithmic structures of the form
\begin{equation}
F^{\rm part}(\Box)\sim \ln\frac{M^2-\Box}{M^2}.
\end{equation}
Note that these partners are still in high-energy domain $-\Box\geq M^2$, but they are subdominant as compared to the leading contribution (\ref{quadratic}) with dimensionless form factors which incorporate the logarithmically running solutions of RG equations. This is because the partners (\ref{CC_partner}) are suppressed by {\em power law} factors $M^4/\Box^2$. Exact form of these formfactors at intermediate scales was derived at one-loop order in \cite{Gorbar-Shapiro_running} for rather generic theory of massive fields by using the heat kernel technique of \cite{CPTI,CPTII}. In IR domain $-\Box\ll M^2$ they are of course expandable in local gradient series reflecting the decoupling phenomenon \cite{Appelquist:1974tg,Gorbar-Shapiro_decoupling,Gorbar-Shapiro_running}.

Similarly, the gravitational constant partner in IR reads as
\begin{equation}
M^2\int d^4x\,\sqrt{g}\Big(R_{\mu\nu}
\frac{F_1(\Box)}{\Box}R^{\mu\nu}+
R\,\frac{F_2(\Box)}{\Box}R\Big),   \label{G_partner}
\end{equation}
which reminds the construction of the nonlocal action for long-distance modifications of gravity theory in \cite{covnonloc,covlongdist}. This differs from the cosmological constant partner by another powers of $M$ and the power of $\Box$ in the denominator.

One should be more careful at this point -- while the case of (\ref{G_partner}) is well defined in asymptotically flat spacetime, the cosmological constant partner (\ref{CC_partner}) is IR divergent for the reasons discussed above. The action of $\tfrac1{\Box^2}$ is not well defined in four dimensions (or, equivalently, $\int d^4x\sqrt{g}(\tfrac1\Box\Re)^2$ is IR divergent), so that the perturbation expansion in the dimension zero sector should be critically reconsidered. To trace the origin of this difficulty note that the first three terms of the cosmological term expansion (\ref{m=0_exp}) are divergent, whereas a similar expansion for the Einstein term becomes well defined only after the subtraction of the Gibbons-Hawking surface term $\int_\infty d^3\sigma^\mu\big(\partial_\mu h-\partial^\nu h_{\mu\nu}\big)$ at the infinity of asymptotically flat spacetime. Due to this subtraction we can write for the integral of the invariant $\Re^{(2)}_1(x)=-R(x)$, weighted in the Einstein action by $\varLambda^{(2)}_1=1/16\pi G$, a legitimate expansion (\ref{invariant}) starting with the quadratic order in $h_{\mu\nu}$,
\begin{align}
&\int d^4x\sqrt{g}(-R)-\int_\infty d^3\sigma^\mu\big(\partial_\mu h-\partial^\nu h_{\mu\nu}\big)\nonumber\\
&\qquad\qquad\quad\quad\; =\sum\limits_{M=2}^\infty
\int d^4x\sqrt{\tilde g}\,I_M^{(2)}(\tilde g,h),\\
&I_2^{(2)}(\tilde g,h)=\!-\tfrac14 h_{\mu\nu}\tilde\Box h^{\mu\nu}\!+\tfrac18 h\tilde\Box h-\tfrac12\big(\tilde\nabla^\nu h_{\mu\nu}\!-\tfrac12\tilde\nabla_\mu\tilde h\big)^2.
\end{align}
Then the above calculational strategy leads to the effective action (\ref{G_partner}) whose tree level IR limit should match low energy physics with the Planck mass cutoff $M^2$ and the form factors $F_1(0)=1$ and $F_2(0)=1/2$. This tree level answer up to $\Re^3$-corrections directly corresponds to the above expression for $I_2^{(2)}(\tilde g,h)$ with $h_{\mu\nu}$ given by Eq. (\ref{h_vs_R}) in terms of the curved space metric $g_{\mu\nu}$ \cite{covnonloc,covlongdist}.

To the best of our knowledge, no such subtraction is known for cosmological term expansion (\ref{m=0_exp}), so that its rigorous treatment is still to be done. It is interesting if new structures can be generated by the regularization of this IR behavior. Apparently, this should be based on the analogue of the Graham-Fefferman construction for asymptotically AdS spaces \cite{Henningson-Skenderis,Haro-Skenderis-Solodukhin} and deserves further studies.

In any case, the UV behavior of both cosmological and gravitational constant partners, which should be not sensitive to IR problems, is determined by curvature squared terms (\ref{quadratic}) with running dimensionless ``couplings''. Their formfactors $F_1(\Box)$ and $F_2(\Box)$ follow from the RG running of the relevant constants $\lambda^{(0)}_2(\mu)$ and $\lambda^{(2)}_2(\mu)$, but the transition $\lambda^{(0,2)}_2(\mu)\to F_{1,2}(\Box)$ is not straightforward and is mediated by Eqs.(\ref{quadratic}) and (\ref{quadratic_G})-(\ref{quadratic_G1}).

\section{Conclusions}
To summarize our notes on conformal anomaly, nonlocal effective action and running scales let us briefly dwell on possible applications of our results and related issues.

As it is clear from the above considerations, the conformal anomaly action is a carrier of the effective rather than fundamental conformal degree of freedom. Either in the nonlocal or the Wess-Zumino form, it is the difference of action functionals of two configurations belonging to the orbit of the conformal group. So unless one of these actions is known the corresponding physical setup is not complete. In this respect, our approach is very different from the works which endow the conformal factor $e^{2\sigma}$ the nature of the fundamental field \cite{Gabadadze} or, for example, ready to sacrifice the fundamental nature of the Higgs boson in favor of 36 fundamental zero dimension scalars $\sigma$ for the sake of a complete eradication of Weyl anomaly and justification of the primordial cosmological perturbations spectra \cite{Boyle-Turok1,Boyle-Turok2}.

The CFT driven cosmology of Sect.\ref{slih} seems to present such an example where the physical setup is complete within a certain approximation scheme. This approximation is associated with the dominance of conformal invariant matter fields over the loop effects of gravity and other types of matter and simultaneously puts the model in the subplanckian domain of energies below the cutoff $M_P/\sqrt\beta$ when the coefficient of the topological conformal anomaly $\beta\gg 1$ \cite{slih_R^2}. To match with the widely accepted bounds on the energy scale of inflation $\sim 10^{-6}M_P$ one needs $\beta\sim 10^{13}$, which cannot be attained by a contribution of low spin conformal fields $\beta=(1/360)(N_0+11N_{1/2}+62N_1)$ unless the numbers $N_s$  of fields of spin $s$ are tremendously high.

On the contrary, this bound can be reached by appealing to the idea of conformal higher spin (CHS) fields \cite{CHS}. A relatively low tower of higher spins will be needed, because a partial contribution of spin $s$ to $\beta$ grows as $s^6$. These partial contributions $\beta_s$ for CHS totally symmetric tensors and Dirac spin-tensors read in terms of $\nu_s$ -- their respective numbers of polarizations (negative for fermions) \cite{Giombietal, Tseytlin13},
    \begin{align}
    &\beta_s=\frac{\nu_s^2(3+14\nu_s)}{720},\, \nu_s=s(s+1),\, s=1,2,3,...\,,\label{boson}\\
    &\beta_s=\frac{\nu_s(12+45\nu_s+14\nu_s^2)}{1440},\nonumber\\
    &\nu_s=-2\Big(s+\tfrac12\Big)^2,\;s=\tfrac12,\tfrac32,\tfrac52,...\,.   \label{fermion}
    \end{align}

The solution of hierarchy problem thus becomes a playground of $1/N$-expansion theory for large number $N$ of conformal
species. Moreover, with the inclusion of CHS fields the status of conformal anomaly essentially changes and becomes similar to that of the chiral anomaly. Chiral anomaly has phenomenological confirmation within chiral symmetry breaking theory, it also has important implications in lepton physics, physics of early Universe, its baryon asymmetry theory, etc. It has a topological nature and is generated in virtue of Adler-Bardeen theorem only at the one-loop level. Local Weyl anomaly also has topological (a-type) contribution \cite{Deser-Schwimmer}, but for low spins it is contributed by all orders of loop expansion. CHS spins, however, have their inverse propagators $\sim\Box^s+...$ and, therefore, for high $s$ are UV finite beyond one loop approximation. So their Weyl anomaly is also exhausted by the one-loop contribution, and there is a hope that their effect in the CFT driven cosmology is nonperturbative. As this effect intrinsically, by a dynamical mechanism of effective equations of motion \cite{slih,slih_R^2,hill-top,SLIH_review}, provides the upper bound on the energy range of inflation $M_P/\sqrt\beta\ll M_P$, this also justifies omission of graviton loops and quantum effects of other (non-conformal) types of matter.

There are, however, serious problems on the road to the realization of this model. To begin with,  CHS fields in curved spacetime are not explicitly known yet, except conformal gravitino with $s=3/2$ and Weyl graviton with $s=2$. Recent progress in generalizing these  models to arbitrary $s$ on the Einstein-space background allowed one to compute their 1-loop Weyl anomaly coefficients (\ref{boson})-(\ref{fermion}) (by indirect AdS/CFT method in \cite{Giombietal} and directly in \cite{Tseytlin13}). This result, however, leaves the issue of unitarity violation caused by inevitable higher derivatives in wave operators of these fields. Moreover, these fields should form a hidden sector not observable at present, which implies the necessity of their eradication in the course of cosmological expansion. What might be useful for this purpose is the idea of renormalization group flow from UV to IR decreasing the value of $\beta$ (the so called $a$-theorem of \cite{Jack-Osborn, KomargodskiSchwimmer, Komargodski}) or Weyl symmetry breaking which would generate masses of CHS fields and thus shorten their massless tower. Finally and most importantly, the fundamental theory of these interacting CHS fields should necessarily be organized within a special higher spin symmetry \cite{Vasiliev00,Vasiliev0}. A complete version of this theory is still missing, not to say about its constructive extension to curved spacetime. Thus, the progress here strongly depends on advancing theory of CHS fields \cite{Tseytlinconf,Segal,Vasiliev,Metsaev,Marnelius}.

The issue of RG running constants $\varLambda$ and $G$ has, as it is shown, a rather unexpected resolution. The manifestation of this UV running actually takes place in the nonlocal formfactors of the quadratic curvature (dimension four) terms rather than in the sector of low dimension operators. This metamorphosis originates from establishing a rather nontrivial scaling operator (\ref{tilde_D}) embracing all powers of the curvature expansion and exploiting a conventional RG assumption that the renormalized theory does not depend on the choice of normalization (or subtraction) point. Then simple, though somewhat tedious, dimensionality considerations lead to this result. Dimension zero and dimension two cosmological and Einstein terms do not run themselves but still contribute to the running of the dimension four terms which can be considered as UV partners of $\varLambda$ and $G$. This metamorphosis of RG running couplings into the formfactors of the curvature squared terms sounds important, because it is the quadratic term in the effective action that mainly determines either the asymptotic freedom of the model or its cutoff beyond which effective field theory breaks down.

In the IR domain these partners, due to the presence of mass scale $M$, also start from the quadratic order in the curvature, but they have essential nonlocality -- of the type $M^4\int d^4x\,\sqrt{g}(\tfrac1\Box\Re)^2$ coming from cosmological constant sector \cite{Donoghue_partners} and of the form $M^2\int d^4x\,\sqrt{g}\Re\tfrac1\Box\Re$ originating from the gravitational constant one. While the latter is well defined in IR limit due to the subtraction from the IR divergent bulk Einstein action of the Gibbons-Hawking surface term \cite{covnonloc,covlongdist}, for the IR cosmological partner \cite{Donoghue_partners} the situation is trickier -- in view of IR divergences it requires additional subtraction procedure. Perhaps even more radical changes will be needed to circumvent this problem like the curvature expansion on top of the homogeneous (dS or AdS) background with nonzero curvature.

Of course, there can be other choices of the running scale $D$ different from (\ref{tilde_D}. Nothing prevents from replacing it, say, with $(\sum_N (-\Box_N)^k)^{1/2k}$ or other combinations of contracted derivatives. However, for curvature squared terms of the action all such choices (satisfying the homogeneity property with respect to derivative rescalings) lead to one and the same operator $\sim(-\Box)^{1/2}$ because for the second order of the curvature expansion all d'Alembertians reduce to the single one, $\Box_1=\Box_2$, in view of integration by parts (\ref{int_by_parts}). The only ambiguity is the choice of the d'Alembertian itself, but it is fixed by the requirement of general covariance. Alterations in the choice of $D$ certainly affect higher orders in the curvature, but the curvature squared part, which is most important for UV asymptotic freedom or determination of the effective field theory cutoff, stays uniquely defined.

Ambiguity in the choice of $D$ can arise in the class of theories which have
a more or less conventional RG running of the gravitational coupling $G$ -- renormalizable Ho\v rava gravity models \cite{Horava,3+1HG}. In these Lorentz symmetry violating models a possible covariant curvature expansion undergoes $(3+1)$-splitting -- the set of basic curvatures includes the extrinsic curvature $K_{ij}$, $i,j=1,2,3$, of spatial slices of constant time $\tau$. The Einstein term of general relativity is replaced by the sum of the kinetic term $\sim(16\pi G)^{-1}\!\int d^4x\,\sqrt{g}K_{ij}^2$ and the potential term built as a polynomial in 3-dimensional curvature and its spatial derivatives. The RG running of $G$ in the kinetic term proceeds as the insertion of the form factor $G^{-1}(D)$ between two factors of $K_{ij}(x)$,
\begin{equation}
\frac1G\!\int d^4x\sqrt{g}\,K_{ij}^2
\to\int d^4x\sqrt{g}\,K_{ij}(x)\frac1{G(D)}K^{ij}(x).
\end{equation}
Thus no tadpole problem for the RG running of $G$ takes place here -- just like in Yang-Mills type theories this occurs  without forming a total derivative structure.

However the relevant scaling operator $D$ of a unit anisotropic scaling dimension, which replaces the spacetime covariant square root of $(-\Box)$, turns out to be ambiguous. Point is that in Lorentz violating models the notion of physical scaling dimension is replaced by the anisotropic one which in $(3+1)$-dimensional Ho\v rava gravity is $-3$ for the time coordinate and $-1$ for spatial coordinates. Correspondingly the dimension $6$ wave operator of the theory is of the second order in time derivatives and of the sixth order in spatial derivatives. Therefore, $D\sim (-\partial_\tau^2-\Delta^3/M^4)^{1/6}$ where $\Delta$ is the spatial covariant Laplacian and $M$ is a physical mass scale parameter. This parameter may be  different in various (scalar and transverse-traceless) sectors of the metric field \cite{3+1HG}, and this is a source of ambiguity in the running scale of Ho\v rava models. Modulo this problem RG running in renormalizable non-projectable Ho\v rava gravity is well defined and in $(3+1)$-dimensional case has a legitimate interpretation of asymptotic freedom \cite{3+1HG}.


\section*{Acknowledgements}
Essential part of this paper was inspired during the workshop ``Quantum Effective Field Theory and Black Hole Tests of Einstein Gravity'' (IFPU, Miramare, Trieste, Italy, September 12-16, 2022), and A.O.B. is very grateful to organizers and participants of this workshop. The authors deeply appreciate the efforts by J.Donoghue, M.Duff, E.Mottola and H.Osborn of critically reading our manuscript. A.O.B. is also grateful for fruitful discussions and correspondence with John Donoghue, Michael Duff, Alexander Kamenshchik, Emil Mottola, Roberto Percacci, Hugh Osborn, Ilya Shapiro,  Kostas Skenderis, Arkady Tseytlin, Alex Vikman, Richard Woodard and especially to G. A. Vilkovisky for long term collaboration on covariant perturbation theory for quantum effective action. This work was supported by the Russian Science Foundation grant No 23-12-00051.

\bibliographystyle{unsrturl} 
\bibliography{Wachowski2307}

\end{document}